\documentclass[preprint,12pt]{elsarticle}

\usepackage[english]{babel}
\usepackage[T1]{fontenc}

\usepackage[letterpaper,top=2cm,bottom=2cm,left=3cm,right=3cm,marginparwidth=1.75cm]{geometry}

\usepackage{amssymb}
\usepackage{amsmath}
\usepackage{graphicx}
\usepackage[colorlinks=true, allcolors=blue]{hyperref}
\usepackage{booktabs}
\usepackage{array}
\usepackage{adjustbox}
\usepackage{caption} 
\usepackage{makecell}
\usepackage{longtable}

\usepackage{chemformula}

\usepackage{epstopdf}

\journal{arXiv}

\begin{document}
    
    \begin{frontmatter}
        
        
        \title{ALMERIA: Boosting pairwise molecular contrasts with scalable methods}
        
        \author[1]{Rafael Mena-Yedra\corref{cor1}}
        \ead{rafael.mena@ual.es}
        
        \author[1]{Juana L. Redondo}
        \ead{jlredondo@ual.es}
        
        \author[2]{Horacio Pérez-Sánchez}
        \ead{hperez@ucam.edu}
        
        \author[1]{Pilar M. Ortigosa}
        \ead{ortigosa@ual.es}
        
        \cortext[cor1]{Corresponding author}
        
            
        \address[1]{Supercomputing - Algorithms Research Group (SAL), University of Almería, Agrifood Campus of International Excellence, ceiA3, 04120, Almería, Spain}
        
            
        \address[2]{Structural Bioinformatics and High Performance Computing Research Group (BIO-HPC), Universidad Católica de Murcia (UCAM), 30107, Murcia, Spain}
        
        \begin{abstract}
    
            Searching for potential active compounds in large databases is a necessary step to reduce time and costs in modern drug discovery pipelines. Such virtual screening methods seek to provide predictions that allow the search space to be narrowed down. Although cheminformatics has made great progress in exploiting the potential of available big data, caution is needed to avoid introducing bias and provide useful predictions with new compounds. In this work, we propose the decision-support tool ALMERIA (Advanced Ligand Multiconformational Exploration with Robust Interpretable Artificial Intelligence) for estimating compound similarities and activity prediction based on pairwise molecular contrasts while considering their conformation variability. The methodology covers the entire pipeline from data preparation to model selection and hyperparameter optimization. It has been implemented using scalable software and methods to exploit large volumes of data ---in the order of several terabytes---, offering a very quick response even for a large batch of queries. The implementation and experiments have been performed in a distributed computer cluster using a benchmark, the public access DUD-E database. In addition to cross-validation, detailed data split criteria have been used to evaluate the models on different data partitions to assess their true generalization ability with new compounds. Experiments show state-of-the-art performance for molecular activity prediction (ROC AUC: $0.99$, $0.96$, $0.87$), proving that the chosen data representation and modeling have good properties to generalize. Molecular conformations ---prediction performance and sensitivity analysis--- have also been evaluated. Finally, an interpretability analysis has been performed using the SHAP method.
            
        \end{abstract}
        
        
        
        
    \end{frontmatter}
    
    \section{Introduction} \label{sec:intro}
    
    Finding a lead compound that can be optimized to result in a drug candidate is an arduous task that entails high costs both in time and monetary terms, mainly because of the vast chemical space. For that reason, computational methods are employed to select a smaller subset of potentially promising compounds for biological testing. This process is known as virtual screening. It comprehends different methods depending on the amount of available information about the compounds in a given database and the biological reactions from previous assays with one of the query compounds or between potentially similar compounds.
    For instance, the simplest approach regarding the amount of utilized prior information is based on the similarity of the compounds. Molecular similarity \cite{maggiora_molecular_2014} could be measured from different perspectives. However, the basic assumption is that structurally similar molecules tend to have similar properties, although this deduction is only sometimes completely evident \cite{martin_structurally_2002}. Moreover, such similarities can be a potential subject of what is known as activity cliff \cite{hu_activity_2020} ---i.e., a small modification to a functional group leads to a sudden change in activity---. As aforementioned, molecular similarity can be measured from different perspectives that include the distance\footnote{Even though any distance metric could be used, in practice, those that are naturally bounded, such as Jaccard/Tanimoto or cosine similarity, are more typically used given their better interpretability.} between molecular descriptors or fingerprints \cite{cereto-massague_molecular_2015}, or alignment-based 3-D similarity, which takes into account rotations and conformations, such as shape similarity \cite{kumar_advances_2018, puertas-martin_optipharm_2019}, among others. The last few years have seen the widespread adoption of deep learning for different problem domains, especially for problems involving unstructured data such as text or images. Deep learning approaches have also been applied for similarity by letting them learn a feature representation in the network latent space. These approaches include neural machine translation \cite{winter_learning_2019, xu_seq2seq_2017} and language models \cite{wang_mol-bert_2021} for string-based representations such as SMILES, variational autoencoders also with SMILES representation \cite{samanta_vae-sim_2020}, or contrastive learning using graph representations \cite{wang_molecular_2022}. The main benefit of these approaches is also their main drawback, as they use unlabelled data and perform data augmentation (atom masking, bond deletion, and subgraph removal) for a given molecule. However, they require labeled data for fine-tuning the last layers in the model in order to achieve a competent performance in downstream tasks.
    
    Another approach is to leverage the available information on known active and inactive compounds to build a predictive model. Input data may also vary from numerical molecular descriptors, the molecular structure in graph notation, and image or string-based representations. Whatever the case, the aim is to find a function mapping from such input space into the output space related to the biological activity between the compounds, for instance. Historically, a reduced number of numerical properties were used in order to quantitatively express their influence on the response variable, which is known as Quantitative-Structure Activity Relationships (QSAR) \cite{cherkasov_qsar_2014} where descriptor selection is fundamental to discard those irrelevant ones beforehand \cite{shahlaei_descriptor_2013, gutierrez-de-teran_modesus_2019}. Later, methods based on machine learning were used for that purpose, being able to handle larger volumes of data \cite{mao_comprehensive_2021} and in a more automated manner. It includes approaches such as convolutional neural networks \cite{stepniewska-dziubinska_development_2018, jimenez_kdeep_2018, ruiz_puentes_pharmanet_2021}, graph networks with attention \cite{cheng_drug-target_2021, hung_qsar_2021, yin_realvs_2021}, or gradient boosting applied with a learning-to-rank procedure \cite{furui_compound_2022}, among others. The main problem with these approaches is precisely the use of annotated data, which is scarce compared to the entire chemical space and expensive to obtain for new pairs of compounds. That is the entire goal of virtual screening, to reduce the potential search space while reducing the risk of omitting promising compounds. Another related problem is derived from the small explored chemical space and the optimization procedures to fit such data that could reward memorization instead of the sought-after generalization for new unseen compounds \cite{wallach_most_2018}.
    
    Lastly, when the 3-D structure of the target molecule is known, structure-based methods such as molecular docking may be applied. However, the space of target molecules with known structures is small compared to the entire space. Additionally, these methods are sensitive to the orientation and torsion of the molecule conformations, but especially to the scoring function \cite{plewczynski_can_2011, shen_beware_2021}.
    
    For a more in depth overview on the literature on virtual screening, the interested reader is referred to a diverse collection of references such as \cite{kimber_deep_2021, deng_artificial_2021, jiang_comprehensive_2021, banegas-luna_review_2018}.
    
    The present work could be framed as a structure-activity relationship (SAR) model but leveraging large-scale high-dimensional data. We put the focus on designing a curated methodology to characterize the molecules in a given database, which comprehends both the modeling of the molecule with different conformations to have a richer 3-D perspective and the generation of nearly 5000 molecular descriptors for every molecule conformation. The choice of using numerical descriptors from the 3-D representations instead of the 3-D representations themselves as images, for instance, is because we consider the former an unbiased representation. It is more easily manageable by algorithms, less noisy, and more likely to have less impact on the time to diagnose and interpret a result. A simple example is that of a group of pixels against certain numerical features of the molecular structure. The work presented here is based on a supervised learning approach to make the most of the activity annotated data, being the ability to generalize with new compounds the ultimate goal of this work. Moreover, this proposal relies on artificial intelligence methods to exploit high-dimensional data instead of over-optimizing based on a specific criterion (e.g., shape). At the same time, efforts have also been put into interpreting the model response and its decision-making ability through explainable AI (XAI).
    
    The document is structured as follows: Section \ref{sec:materialsmethods} describes the proposed materials and methods using a generic and modular architecture, while the specific implementation details, as well as the obtained results, are discussed in Section \ref{sec:results}. In Section \ref{sec:interpertability}, an interpretability analysis is performed on the obtained model to assess its decision-making ability. Some performance measurements ---both in CPU and GPU--- are given in Section \ref{sec:perf-measurement}. Finally, in Section \ref{sec:conclusion}, conclusions and potential future work are drawn.
    
    \section{Materials and methods} \label{sec:materialsmethods}
    
\begin{figure}
        \centering
        \includegraphics[width=0.99\textwidth]{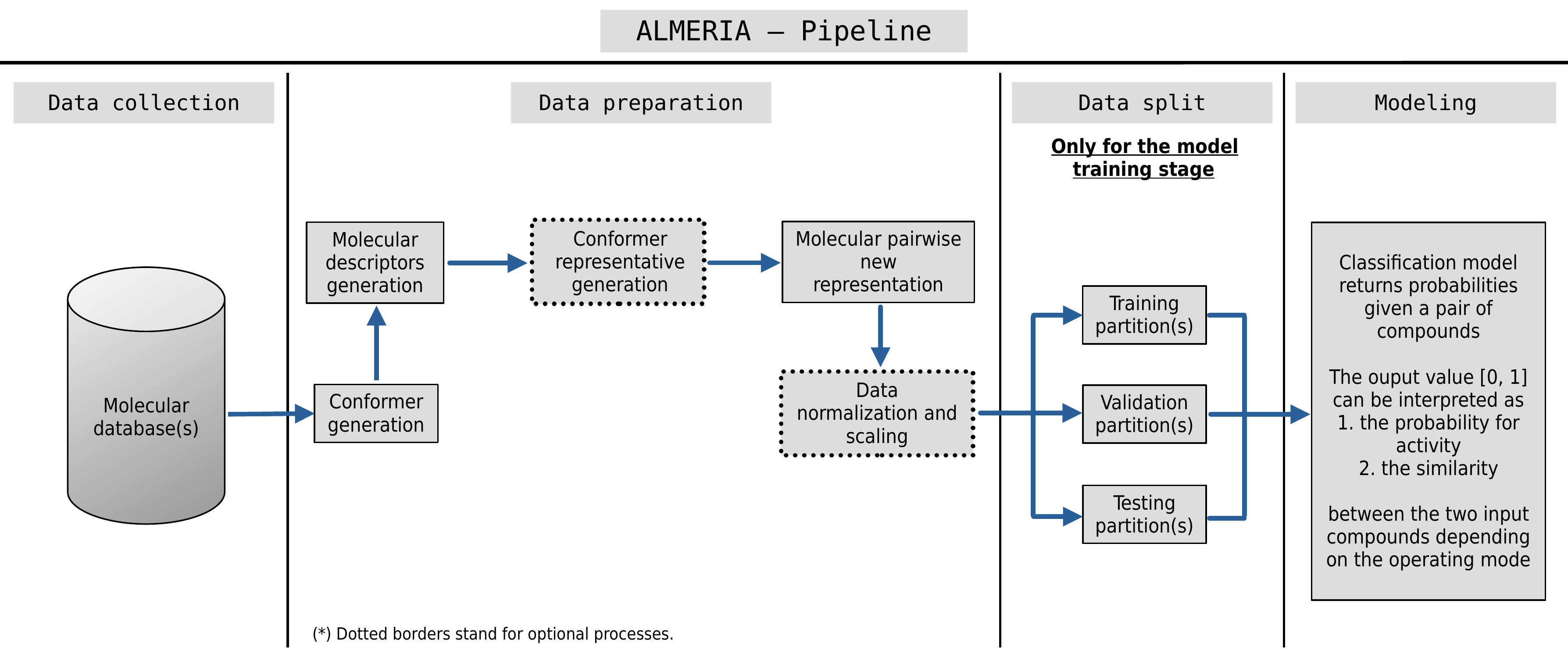}
        \caption{\label{fig:pipeline}Overall scheme for ALMERIA using the proposed materials and methods.}
    \end{figure}

    The general scheme showing the flow of information using the proposed materials and methods can be seen in Figure \ref{fig:pipeline}. The boxes in the diagram intentionally show generic names for the proposed materials and methods, intending to modularise the methodology to reflect the flexibility to replace specific portions of the proposal. In this section, the functionality covered in each part will be described briefly and from a fundamental point of view. However, it will be in Section \ref{sec:experiments-setup} where more details on the specific implementation for this work will be given.
    
    \subsection{Data collection}
    
    Any molecular database could be used to feed the pipeline shown in Figure \ref{fig:pipeline}. The only requisite for the model training stage is to have a certain number of active compounds that can consist of a labeled dataset where a binary label either active or not may suffice. Of course, this requirement does not apply to the prediction mode as it is the final aim to predict the potential biological activity between new molecules. ALMERIA has therefore been designed to make the most of the already activity-labeled data; it is not necessary to have the entire database(s) already labeled with the biological activity, but a subset may suffice. However, as always with data-driven models, the more and better quality data, the more likely it is that a good, generalizable model will be obtained.
    
    In contrast, the size and dimensionality of the molecular database(s) are not restricted because the methodology and implementation details have been chosen carefully to be scalable. These include using big data-oriented software such as Dask \cite{dask_development_team_dask_2016}, allowing the entire pipeline operations to be executed in parallel and distributed over a cluster of computers. 
    
    As the molecules in the database are given in a rigid state, we use the software OpenEye Scientific Omega \cite{hawkins_conformer_2010} to generate a set of 3-D molecular conformations for every compound in the database. That way, we have a broader representation of a molecule's different conformations. However, care must be taken as the number of conformations for a given molecule may suffer from a combinatorial explosion. If each torsion angle is rotated in increments of $\theta$ degrees for a molecule with $N$ rotatable bonds, then the total number of conformations would be $(360^{\circ} / \theta)^{N}$. 
    
    Next, we use these molecular conformations to generate a large set of $4\,885$ numerical descriptors using the software Dragon \cite{mauri_dragon_2006}. The list of descriptors includes the simplest atom types, functional groups and fragment counts, topological and geometrical descriptors, 3-D descriptors, and several properties estimation (such as logP) and drug-like and lead-like alerts (such as Lipinski's alert). Table \ref{tab:dragon_descriptors_blocks} shows the number of descriptors grouped by logical blocks. The choice of representing the molecular conformations using numerical descriptors instead of alternative based on unstructured data such as 2-D/3-D images or string-based is ---in our view--- mainly due to an unbiased and less noisy representation, a potentially more interpretable decision-making analysis and more efficient management by algorithms. Thus, through this work, the data has been arranged in a tabular form where columns correspond to features ---molecular descriptors--- and rows correspond to data samples ---molecules or pairs of molecules as will be described in Section \ref{sec:data-preparation}---.
    
    \begin{table}[]
        \centering
        \caption{Number of molecular descriptors grouped by logical block using Dragon software.}
        \label{tab:dragon_descriptors_blocks}
        \begin{tabular}{@{}lc@{}}
            \toprule
            \textbf{Block name}         & \multicolumn{1}{l}{\textbf{No. of descriptors}} \\ \midrule
            Constitutional descriptors  & 43                                              \\
            Ring descriptors            & 32                                              \\
            Topological indices         & 75                                              \\
            Walk and path counts        & 46                                              \\
            Connectivity indices        & 37                                              \\
            Information indices         & 48                                              \\
            2D matrix-based descriptors & 550                                             \\
            2D autocorrelations         & 213                                             \\
            Burden eigenvalues          & 96                                              \\
            P-VSA-like descriptors      & 45                                              \\
            ETA indices                 & 23                                              \\
            Edge adjacency indices      & 324                                             \\
            Geometrical descriptors     & 38                                              \\
            3D matrix-based descriptors & 90                                              \\
            3D autocorrelations         & 80                                              \\
            RDF descriptors             & 210                                             \\
            3D-MoRSE descriptors        & 224                                             \\
            WHIM descriptors            & 114                                             \\
            GETAWAY descriptors         & 273                                             \\
            Randic molecular profiles   & 41                                              \\
            Functional group counts    & 154                                             \\
            Atom-centered fragments      & 115                                             \\
            Atom-type E-state indices   & 170                                             \\
            CATS 2D                     & 150                                             \\
            2D Atom Pairs               & $1\,596$                                        \\
            3D Atom Pairs               & 36                                              \\
            Charge descriptors          & 15                                              \\
            Molecular properties        & 20                                              \\
            Drug-like indices           & 27                                              \\ \bottomrule
        \end{tabular}
    \end{table}
    
    \subsection{Data preparation} \label{sec:data-preparation}
    
    Once data is collected, we have decided not to perform any major data transformation or dimensionality reduction along the descriptors axis to avoid harming the interpretability of the model. However, we opted for performing two preprocessing steps to both favor the method generalization and make the process more efficient:
    
    \begin{enumerate}
        \item The set of conformations for every compound is reduced to a single representative sample by averaging their descriptor values ---i.e., grouping by molecule and then averaging their descriptors column-wise---. Thus, this representative sample is built considering the different conformations the molecule may adopt. As a side effect, this improves the efficiency of model building. The first benefit sought with this step is fairer when guiding the model optimization and evaluating its performance, as activity data is usually molecule-wise labeled. However, conformation generation may imply that certain molecules are overrepresented in comparison to others, which may bias both the optimization and the evaluation metrics, i.e., risk of frequency bias. In any case, this step is optional for ALMERIA within the proposed methodology shown in Figure \ref{fig:pipeline}, and all conformations could be used for modeling. However, since we have included this step in the current work and know that it can be controversial outside the field of machine learning, we have included Subsection \ref{sec:sensitivity-analysis} within the experimentation that analyzes its impact on the set of generated conformations.
        \item Instead of building a separate model for every target, as often found in literature, we opt for building a single model that considers the specific contrast between compounds that correlates with biological activity. We reach this by performing the absolute difference on the descriptors for every pair of target and ligand molecules. It aims to improve the generalization performance with compounds not yet seen during the model fitting while making the ALMERIA methodology more efficient with a single model without sacrificing interpretability.
    \end{enumerate}
    
    \subsection{Data split}
    
    In addition to the previous data modeling choices, we have also been careful during the model-building stage to make the most efficient use of the available data to maximize the generalization. Given that we are adopting a supervised learning approach, we have considered that such labeled data is expensive to obtain. This consideration is important, as it is well known that poorly trained machine learning models can easily over-fit the training data and perform much worse with new and potentially different molecules. For this reason, based on activity value, we perform a stratified $K=10$ cross-validation (CV) to select the model configuration that better generalizes during a hyperparameter optimization (HPO) process. Additionally, another data partition (test data set) was held to validate the out-of-sample partition, i.e., the ability to generalize with unseen compounds. Moreover, because we are using a single sample for every compound during the whole process (training, validation, and testing), it implies the process is always designed to validate every k-th fold and test on compounds not seen during training. It forces the method to find patterns that truly generalize between compounds.
    
    \subsection{Modeling}
    
    The basic modeling choice has been to use a classification model. That is, given an input data set $X$, the model will return a probability $y$ that quantifies its confidence in the potential activity between a pair of molecules:
    
    $$ f : X \rightarrow y $$
    
    Any model that fulfills this basic criterium may be included within the ALMERIA methodology shown in Figure \ref{fig:pipeline}. This fact includes our main proposal gradient boosting (according to the most important needs we found) as well as a set of competitive baselines used to benchmark the performance against the main proposal, thus validating its best suitability. In the following subsections, we will describe several examples that could be used as the machine learning algorithmic $f(X)$ within the ALMERIA methodology.
    
    \subsubsection{The main proposal: Gradient boosting} \label{sec:gradient-boosting}
    
    In order to choose the most appropriate data-driven modeling approach, the main characteristics that underlie the problem, as well as the potential features the solution should offer, have been considered:
    
    \begin{itemize}
        \item Complex and non-linear mapping from feature to output space.
        \item Structured input data in a high-dimensional space and big volume. It requires an efficient approach that can also scale to easily accommodate increasing volumes of data, possibly leveraging more hardware in a distributed environment.
        \item Expensive but valuable annotated data, thus leveraging a supervised learning approach to get the most out of prior efforts.
        \item The importance of having annotated data also lies in being able to assess the performance of the model on new out-of-sample molecules not seen during model fitting. For this reason, having a high-capacity model is as important as having tools to avoid memorization and overfit, a potential pitfall in the field \cite{wallach_most_2018}.
        \item Boosting the ease of interpreting both the model output and which factors influence the most on its decision.
    \end{itemize}
    
    These have shaped the decision towards a gradient boosting implementation as these have been used widely in industry and popular competitions among data scientists and machine learning practitioners and are recognized for dealing properly with the needs mentioned above. More specifically, the open-source software library XGBoost \cite{github_xgboost} was chosen as it provides an optimized distributed gradient boosting framework designed to be highly efficient and flexible. This library perfectly suits the environment where the present work has been developed using a high-performance computing environment.
    
    In machine learning literature, gradient boosting can be seen as the follow-up in the \textit{natural evolution} and modeling refinement from decision trees or CART (classification and regression trees) \cite{breiman_classification_1984}, and random forests \cite{breiman_random_2001} which handle an ensemble, by using \textit{bagging}, of the former to reduce the variance and overfitting. Models relying on decision trees as a base learner have been fruitfully applied ---if properly controlling their complexity--- to data modeling problems with non-linear decision boundaries. Boosting originates from the idea of iteratively adding weak learners that improve the previous error, thus generating a collectively boosted strong model. Gradient boosting \cite{breiman_arcing_1997, friedman_greedy_2001} makes this boosting setting very efficient by using the gradient of the error from the corresponding objective function to guide the ensemble construction.
    
    In this work, the chosen training loss has been the well-known binary cross entropy: $ L(\theta) = \sum_i[ y_i\ln (1+e^{-\hat{y}_i}) + (1-y_i)\ln (1+e^{\hat{y}_i})] $. The motivation behind this choice is that the resulting output value is a probability $y \in [0, 1]$, which is straightforward to understand for end-users as a proxy for the similarity or affinity between two chemical compounds. Actually, the objective function is composed both of the training loss mentioned above that assesses the correct mapping $ f: X \rightarrow y $, and the regularization term $ \Omega(\theta) $ that serves as a control for the complexity of the model to avoid overfitting to the training data $ \text{obj}(\theta) = L(\theta) + \Omega(\theta) $.
    
    The choice of XGBoost as the gradient boosting implementation is based on its design for large-scale machine-learning \cite{chen_xgboost_2016}, including several optimizations such as building trees in a parallel way instead of sequentially like the original gradient boosting or data sketching with histograms, among others.
    
    XGBoost has associated several parameters that define its inner working when building the model allowing it to adjust its complexity properly to the problem. The best way to adjust them is through a hyperparameter optimization (HPO) process based on the data partitioning schema using cross-validation as explained in Section \ref{sec:data-preparation}. The HPO has been carried out using the state-of-the-art framework Optuna \cite{akiba_optuna_2019}. 
    
    \subsubsection{Baselines} \label{sec:baselines}
    
    To contextualize the resulting predictive performance from the modeling proposal, we have included a diverse set of baselines that comprehend different machine learning algorithmic strategies. This inclusion also highlights the modularity of the ALMERIA methodology shown in Figure \ref{fig:pipeline}.
    
    The underlying data preprocessing and preparation are the same as described in earlier sections. The only difference is in the performed data preprocessing because of the limitation for some of the baseline models to deal with: missing data in the input features, columns with almost zero variance during model fitting, as well as applying a Z-score normalization as a preprocessing layer in order to handle input features at similar scales. Thus, when necessary, according to the model requirements, the following additional preprocessing steps have been applied: 
    
    \begin{enumerate}
        \item Replace missing numerical data entries with a simple imputation strategy using the mean value from the corresponding feature.
        \item Drop features columns whose variance is almost zero, i.e., constant values.
        \item Apply Z-score normalization to transform the different features into the same scale with 0 mean and 1 standard deviation. Statistics used to apply the normalization are calculated from the training data partition to avoid data leakage.
    \end{enumerate} 
    
    The following baseline models have been selected: logistic regression, support vector machines (SVM) using an ensemble voting approach, random forests, and a deep neural network with a dense architecture for classification. All these models have also followed a hyperparameter optimization process before fully training the final model to be evaluated on the testing data partitions. As for the specific software implementation: we have used \textit{XGBoost} for gradient boosting, \textit{Keras/Tensorflow} for the deep learning approach, and \textit{scikit-learn} for the rest of baselines as software packages.
    
    \paragraph{Logistic regression:}
    
    Logistic regression \cite[Chapter~4]{bishop_pattern_2006} has been included as representative of a linear model. Despite being one of the simplest machine learning models, that is precisely its core strength as it has less room for overfitting, allowing it often to generalize well.
    
    Among the assumptions and requirements for a logistic regression model:
    \begin{itemize}
        \item Absence of certain multicollinearity among the independent variables ---molecular descriptors in our case--- cannot be guaranteed, and thus such assumption could be violated. Still, perfectly collinear columns are odd to be found.
        \item Columns with almost zero variance ---i.e., constant value--- have been dropped to avoid perfectly collinear columns and be able to solve the model.
        \item Missing data entries have been filled with the corresponding average value.
        \item Given that regularization has been considered for the model fitting, Z-score normalization has been applied to data ---using statistics from the training data partition---.
    \end{itemize}
    
    \paragraph{SVM - ensemble:}
    
    Support vector machines (SVM) \cite[Chapter~7]{bishop_pattern_2006} have been chosen as another baseline model. In this case, SVM may learn non-linear decision boundaries as the selected underlying kernels ---radial basis function, polynomial or sigmoid--- allow it. Moreover, an ensemble strategy has been considered instead of fitting a single SVM for the entire dataset. It means that several SVM with the same hyperparameters configuration are learned across different data subsamples. Then, the final output is agreed upon based on the $argmax$ of the sums of the predicted probabilities.
    
    Among the requirements for an SVM model:
    \begin{itemize}
        \item Columns with almost zero variance ---i.e., constant value--- have been dropped to avoid perfectly collinear columns.
        \item Missing data entries have been filled with the corresponding average value.
        \item Given the regularization considered for the model fitting, Z-score normalization has been applied to data by using statistics from the training data partition.
    \end{itemize}
    
    %
    %
    %
    
    \paragraph{Random forests:}
    
    The random forests \cite{breiman_random_2001} model has been included as another baseline, as it shares some of its underlying algorithmic principles with the main modeling proposal based on gradient boosting.
    
    Among the requirements for a random forest model:
    \begin{itemize}
        \item Columns with almost zero variance ---i.e., constant value--- have been dropped to avoid perfectly collinear columns.
        \item Missing data entries have been filled with the corresponding average value.
    \end{itemize}
    
    \paragraph{Deep neural network:}
    
    Despite the successful application of deep learning for perceptual tasks and unstructured data, its success in tasks with tabular data has been more modest compared to other approaches. Still, there have been some clever approaches to deal with tabular data using the recently ubiquitous Transformer architecture \cite{huang_tabtransformer_2020}. However, its potential is focused on categorical input data to provide them with attention mechanisms. Conversely, for the problem at hand, all the input data ---molecular descriptors--- is numerical data.
    
    Therefore, the architecture used here is based on a feed-forwarded structure. Besides the input layer and the sigmoid output layer for classification, two blocks can be differentiated in the central part of the network structure. The first part is composed of $N$ consecutive feed-forwarded connected sub-blocks where each one is composed of a Layer Normalization layer, a Dense layer, and a Dropout layer ---this last one could be omitted according to the HPO process---. Every $n^{th}$ sub-block has a total of $16.7$ millions of parameters. Then, the second part is similar to the first one. However, there is a single Layer Normalization at the beginning, and each of the $M$ consecutive feed-forwarded sub-blocks ---composed of a Dense layer and an optional Dropout layer--- gradually decreases its number of hidden nodes. The number of parameters for this second part may vary from $0$ to $400$ million according to the number of $M$ sub-blocks.
    
    Among the requirements for this deep learning model:
    \begin{itemize}
        \item Columns with almost zero variance ---i.e., constant value--- have been dropped to avoid perfectly collinear columns.
        \item Missing data entries have been filled with the corresponding average value.
    \end{itemize}
    
    Moreover, regarding the Z-score normalization, two versions with the same hyperparameter setup have been optimized to assess the input data normalization effect. For example, it is known that normalizing the input data for a deep learning model favors the convergence properties of the optimization algorithm.
    
    \section{Results and discussion} \label{sec:results}
    
    \subsection{Experiment setup} \label{sec:experiments-setup}
    
    This section will provide details on the specific implementation that has been made in each module of Section \ref{sec:materialsmethods}.
    
    As for the molecular database used in this work, the Directory of Useful Decoys - Enhanced (DUD-E) \cite{mysinger_directory_2012} has been used to assess and validate the modeling proposal in this work. It is a target-ligand public database well-known in the field and is often used to validate and quantify the performance of a virtual screening methodology. Overall, it contains 102 target proteins and 22\,886 active compounds ---an average of 224 ligands per target---. In addition, there are 50 decoys ---or inactive compounds--- for each active compound, having similar 1-D Physico-chemical properties to remove bias (e.g., molecular weight, calculated LogP) but different 2-D topology to be likely non-binders. The total number of compounds exceeds 1.4 million (22\,886 actives, and 1\,411\,214 decoys).
    
    Since the database contains rigid molecules, up to 100 3-D molecular conformations have been generated for every compound to consider its flexibility. Molecular descriptors are extracted from this extended database with conformations, generating new representations per pair of molecules. Then, for the main modeling proposal ---gradient boosting--- there is no more data preprocessing. For the rest of the comparative baselines, the preprocessing details are specified in Section \ref{sec:baselines}.
    
    Next, for the model training stage, the data is split into three parts: one for training that will be used with cross-validation using $K=10$ folds for hyperparameter optimization. The other two partitions not used during training will be used for testing with two different aims.
    
    More specifically, we have proceeded with the following data partitioning schema using all the 102 target proteins and their associated ligand compounds ---either active or decoy---. The same data partitioning schema, as well as random seeds, have been used for all the experiments run and baseline models:
    
    \begin{itemize}
        \item \textbf{Data partition A}: 96 out of the 102 target proteins along with their associated ligand compounds. The list of target proteins for this partition is: \texttt{ACES}, \texttt{ADA}, \texttt{ADRB1}, \texttt{ADRB2}, \texttt{AKT2}, \texttt{ALDR}, \texttt{AMPC}, \texttt{AOFB}, \texttt{BACE1}, \texttt{BRAF}, \texttt{CAH2}, \texttt{CASP3}, \texttt{CDK2}, \texttt{COMT}, \texttt{CP2C9}, \texttt{CP3A4}, \texttt{CSF1R}, \texttt{CXCR4}, \texttt{DEF}, \texttt{DHI1}, \texttt{DPP4}, \texttt{DRD3}, \texttt{DYR}, \texttt{EGFR}, \texttt{ESR1}, \texttt{ESR2}, \texttt{FA10}, \texttt{FA7}, \texttt{FABP4}, \texttt{FAK1}, \texttt{FGFR1}, \texttt{FKB1A}, \texttt{FNTA}, \texttt{FPPS}, \texttt{GCR}, \texttt{GLCM}, \texttt{GRIA2}, \texttt{GRIK1}, \texttt{HDAC2}, \texttt{HDAC8}, \texttt{HIVINT}, \texttt{HIVPR}, \texttt{HIVRT}, \texttt{HMDH}, \texttt{HS90A}, \texttt{HXK4}, \texttt{IGF1R}, \texttt{INHA}, \texttt{ITAL}, \texttt{JAK2}, \texttt{KIF11}, \texttt{KIT}, \texttt{KITH}, \texttt{KPCB}, \texttt{LCK}, \texttt{LKHA4}, \texttt{MAPK2}, \texttt{MCR}, \texttt{MET}, \texttt{MK01}, \texttt{MK10}, \texttt{MK14}, \texttt{MMP13}, \texttt{MP2K1}, \texttt{NOS1}, \texttt{NRAM}, \texttt{PA2GA}, \texttt{PARP1}, \texttt{PDE5A}, \texttt{PGH1}, \texttt{PGH2}, \texttt{PLK1}, \texttt{PNPH}, \texttt{PPARA}, \texttt{PPARD}, \texttt{PPARG}, \texttt{PRGR}, \texttt{PTN1}, \texttt{PUR2}, \texttt{PYGM}, \texttt{PYRD}, \texttt{RENI}, \texttt{ROCK1}, \texttt{RXRA}, \texttt{SAHH}, \texttt{SRC}, \texttt{TGFR1}, \texttt{THB}, \texttt{THRB}, \texttt{TRY1}, \texttt{TRYB1}, \texttt{TYSY}, \texttt{UROK}, \texttt{VGFR2}, \texttt{WEE1}, \texttt{XIAP}.
        \begin{itemize} 
            \item \textbf{Data partition A.1}: $70\%$ from data partition A has been used to train the models using a $K=10$ cross-validation setting.
            \item \textbf{Data partition A.2}: $30\%$ from data partition A has been considered for testing the model's accuracy after they have been trained with partition A.1. This sub-partitioning implies that target proteins from partition A have been mixed among partitions A.1 and A.2. Therefore they could be present in both or just in one of them, but every ligand compound is either in partition A.1 or A.2. This allows assessing the model with new ligands not seen before during training.
        \end{itemize}
        \item \textbf{Data partition B}: 6 out of the 102 target proteins and their associated ligand compounds. The list of target proteins for this partition is: \texttt{AKT1}, \texttt{ACE}, \texttt{AA2AR}, \texttt{ABL1}, \texttt{ANDR}, \texttt{ADA17}. This selection has been made by hand to cope with one target compound per DUD-E subset: Diverse, Dud38, GPCR, Kinase, Nuclear, and Protease. This partition allows the assessment of the model with new targets and ligand compounds not seen before during training.
    \end{itemize}
    
    Finally, all the methods and experiments included in this work have been implemented in the distributed computer cluster managed by the Supercomputing and Algorithms research group at the University of Almeria \cite{university_of_almeria_supercomputing_nodate}. Briefly, the cluster has a total of 33 nodes with 74 CPUs + 15 GPUs (1\,380 cores, 9 TB of RAM, and 25 TB of solid-state storage) that are distributed over an Infiniband network as follows:
    
    \begin{itemize}
        \item Front-end: Bullx R423E3i. 2 Intel Xeon E5 2620 2 GHz (12 cores) and 64 GB RAM. RAID disk with 16 TB.
        \item 8x Bull Sequana X440-A5: 2 AMD EPYC Rome 7642 (48 cores) and 512 GB RAM. 240 GB SSD.
        \item 2x Bull Sequana X410-A5: 2 AMD EPYC Rome 7302 (16 cores) and 512 GB RAM. 240 GB SSD.
        \begin{itemize}
            \item 4x GPUs NVIDIA Tesla V100 with 32GB HBM2, 5\,120 CUDA cores and 640 Tensor cores.
        \end{itemize}
        \item 2x Bullx R421-E4: 2 Intel Xeon E5 2620v3 (12 cores) and 64 GB RAM. 1 TB HDD.
        \begin{itemize}
            \item 2x NVIDIA K80: 2 Kepler GK210 with 24 GB GDDR5 and 4\,992 cores CUDA.
            \item AMD ATI SAPPHIRE FIRE PRO S9100: 2\,560 stream processors and 12 GB GDDR5.
        \end{itemize}
        \item Bullion S8: 8 Intel Xeon E7 8860v3 (16 cores) and 2.3 TB RAM. 2x 300 GB SAS.
        \item 18x Bullx R424-E3: 2 Intel Xeon E5 2650 (16 cores) and 64 GB RAM. 128 GB SSD.
        \item 2x Bullx R424-E3: 2 Intel Xeon E5 2650v2 (16 cores) and 128 GB RAM. 1 TB HDD.
        \item 2x NextIO 2070.
        \begin{itemize}
            \item 4x GPUs Tesla M2070 (1\,792 cores).
        \end{itemize}
    \end{itemize}
    
    \subsection{Hyperparameter optimization}
    
    The training partition (A.1) has been used for the hyperparameter optimization (HPO) process for all the models (gradient boosting and the rest of the baselines) using 100 trials per HPO process. The HPO has been carried out using the state-of-the-art framework Optuna \cite{akiba_optuna_2019}. Then, details about the HPO parameters and results will be given for every model.
    
    \paragraph{Gradient boosting}
    
    The search space has been defined as shown in Table \ref{tab:hpo_xgb_searchspace}, and the best found hyperparameter set is shown in Table \ref{tab:hpo_xgb_best}.
    
    \begin{table}[]
        \centering
        \caption{Search space used during the hyperparameter optimization (HPO) process for fine-tuning the XGBoost model.}
        \label{tab:hpo_xgb_searchspace}
        \begin{tabular}{@{}llp{7cm}@{}}
            \toprule
            \textbf{Hyperparameter} & \textbf{Search space}                                  & \textbf{Description} \\ \toprule
            Grow policy             & $x \in \{\textrm{`depthwise'}, \textrm{`lossguide'}\}$ & Split either at nodes closest to the root or at nodes with highest loss change. \\ \midrule
            No. of estimators       & $1 \leq x \leq 1000$                                   & Number of boosting iterations. \\ \midrule
            Learning rate           & $x = 0.05$                                             & Step size shrinkage used in update to prevent overfitting. \\ \midrule
            Maximum depth           & $x \in \{3, 5, 7, 9, 11, 13\}$                         & Maximum depth per boosting iteration (tree). \\ \midrule
            Min. child weight       & $x \in \{1, 5, 10\}$                                   & Minimum sum of instance weight (Hessian) needed in a child. \\ \midrule
            Alpha                   & $x \in \{0, 0.5, 1, 2, 5\}$                            & L1 regularization term on weights. \\ \midrule
            Lambda                  & $x \in \{0, 0.5, 1, 2, 5\}$                            & L2 regularization term on weights. \\ \midrule
            Min. split loss         & $x \in \{0, 0.5, 1, 2, 5\}$                            & Minimum loss reduction required to make a further partition on a leaf node of the tree. \\ \midrule
            Max. delta step         & $x \in \{0, 0.5, 1, 2, 5, 10\}$                        & Values higher than zero make the update step more conservative. \\ \midrule
            Subsample               & $0.2 \leq x \leq 1$                                    & Subsample ratio of the training instances in every boosting iteration. \\ \midrule
            Columns subsample       & $0.2 \leq x \leq 1$                                    & Subsample ratio of columns in every boosting iteration. \\ \bottomrule
        \end{tabular}
    \end{table}
    
    \begin{table}[]
        \centering
        \caption{Best hyperparameter set found after the HPO process for the XGBoost model.}
        \label{tab:hpo_xgb_best}
        \begin{tabular}{@{}ll@{}}
            \toprule
            \textbf{Hyperparameter} & \textbf{Value}              \\ \toprule
            Grow policy             & \textrm{`lossguide'}        \\ \midrule
            No. of estimators       & $992$                       \\ \midrule
            Learning rate           & $0.05$                      \\ \midrule
            Maximum depth           & $9$                         \\ \midrule
            Min. child weight       & $5$                         \\ \midrule
            Alpha                   & $0.5$                       \\ \midrule
            Lambda                  & $2.0$                       \\ \midrule
            Min. split loss         & $0.5$                       \\ \midrule
            Max. delta step         & $5.0$                       \\ \midrule
            Subsample               & $1.0$                       \\ \midrule
            Columns subsample       & $0.4$                       \\ \bottomrule
        \end{tabular}
    \end{table}
    
    \paragraph{Logistic regression}
    
    Table \ref{tab:hpo_lr_searchspace} shows the search space defined for the hyperparameter optimization process of the logistic regression model, and Table \ref{tab:hpo_lr_best}  shows the best-found hyperparameter set.
    
    \begin{table}[]
        \centering
        \caption{Search space used during the hyperparameter optimization (HPO) process for fine-tuning the logistic regression model.}
        \label{tab:hpo_lr_searchspace}
        \begin{tabular}{@{}lp{4cm}p{7cm}@{}}
            \toprule
            \textbf{Hyperparameter} & \textbf{Search space}                                     & \textbf{Description}                                                                  \\ \toprule
            Penalty                 & $x \in \{L_{1}, L_{2}\}$                                  & Norm of the penalty for the bias/variance trade-off.                                  \\ \midrule
            C                       & $x \in \{0.001, 0.01, 0.1,$ $0.5, 1.0, 2.0, 5.0, 10.0\}$    & Inverse of regularization strength; smaller values specify stronger regularization.   \\ \bottomrule
        \end{tabular}
    \end{table}
    
    \begin{table}[]
        \centering
        \caption{Best hyperparameter set found after the HPO process for the logistic regression model.}
        \label{tab:hpo_lr_best}
        \begin{tabular}{@{}ll@{}}
            \toprule
            \textbf{Hyperparameter} & \textbf{Value} \\ \toprule
            Penalty                 & $L_{2}$        \\ \midrule
            C                       & $0.001$        \\ \bottomrule
        \end{tabular}
    \end{table}
    
    \paragraph{SVM - ensemble}
    
    Table \ref{tab:hpo_svm_searchspace} shows the search space for the hyperparameter optimization process of the SVM model, and Table \ref{tab:hpo_svm_best} indicates the best-found hyperparameter set.
    
    \begin{table}[]
        \centering
        \caption{Search space used during the hyperparameter optimization (HPO) process for fine-tuning the SVM model.}
        \label{tab:hpo_svm_searchspace}
        \begin{tabular}{@{}lp{4cm}p{7cm}@{}}
            \toprule
            \textbf{Hyperparameter} & \textbf{Search space}                                     & \textbf{Description}                                                                  \\ \toprule
            Kernel                  & $x \in \{\textrm{`radial-basis'},$ $\textrm{`polynomial'},$ 
            $\textrm{`sigmoid'}\}$    & Kernel type for every SVM instance.                                  \\ \midrule
            C                       & $x \in \{0.001, 0.01, 0.1$ $0.5, 1.0, 2.0, 5.0, 10.0\}$    & Inverse of regularization strength; smaller values specify stronger $L_{2}$ regularization.   \\ \bottomrule
        \end{tabular}
    \end{table}
    
    \begin{table}[]
        \centering
        \caption{Best hyperparameter set found after the HPO process for the SVM model.}
        \label{tab:hpo_svm_best}
        \begin{tabular}{@{}ll@{}}
            \toprule
            \textbf{Hyperparameter} & \textbf{Value} \\ \toprule
            Kernel                 & $\textrm{`radial-basis'}$        \\ \midrule
            C                       & $0.5$        \\ \bottomrule
        \end{tabular}
    \end{table}
    
    \paragraph{Random forests}
    
    Table \ref{tab:hpo_rf_searchspace} shows the search space for the hyperparameter optimization process of the random forest model, and Table \ref{tab:hpo_rf_best} indicates the best-found hyperparameter set.
    
    \begin{table}[]
        \centering
        \caption{Search space used during the hyperparameter optimization (HPO) process for fine-tuning the random forest model.}
        \label{tab:hpo_rf_searchspace}
        \begin{tabular}{@{}lp{4cm}p{7cm}@{}}
            \toprule
            \textbf{Hyperparameter} & \textbf{Search space}                                     & \textbf{Description}                                                                  \\ \toprule
            Split criterion & $x \in \{\textrm{`gini'},$ $\textrm{`entropy'},$ 
            $\textrm{`log-loss'}\}$    & The function to measure the quality of a split.                                  \\ \midrule
            No. of estimators (trees) & $x \in \{10, 100, 1000, 10000\}$    & The number of trees in the forest.   \\ \bottomrule
            Min. samples split & $x \in \{2, 4, 16, 256, 1024\}$    & The minimum number of samples required to split an internal node.   \\ \bottomrule
        \end{tabular}
    \end{table}
    
    \begin{table}[]
        \centering
        \caption{Best hyperparameter set found after the HPO process for the random forest model.}
        \label{tab:hpo_rf_best}
        \begin{tabular}{@{}ll@{}}
            \toprule
            \textbf{Hyperparameter} & \textbf{Value} \\ \toprule
            Split criterion                 & $\textrm{`entropy'}$        \\ \midrule
            No. of estimators (trees)   & $10000$        \\ \bottomrule
            Min. samples split   & $4$        \\ \bottomrule
        \end{tabular}
    \end{table}
    
    \paragraph{Deep neural network}
    
    Table \ref{tab:hpo_nn_searchspace} shows the search space for the hyperparameter optimization process of the deep learning model, Table \ref{tab:hpo_nn_best_Z} indicates the best-found hyperparameter set for the version with Z-score normalization, and Table \ref{tab:hpo_nn_best_noZ} for the version without input data normalization.
    
    \begin{table}[]
        \centering
        \caption{Search space used during the hyperparameter optimization (HPO) process for fine-tuning the deep learning model.}
        \label{tab:hpo_nn_searchspace}
        \begin{tabular}{@{}lp{4cm}p{7cm}@{}}
            \toprule
            \textbf{Hyperparameter} & \textbf{Search space} & \textbf{Description} \\ \toprule
            Batch size & $x \in \{256, 512, 1024,$ $2048, 4096\}$ & Batch size for every gradient update. \\ \midrule
            Optimization algorithm & $\textrm{`Adam'}$ & --- \\ \midrule
            Initial learning rate & $x \in \{1e-1, 1e-2,$ $1e-3, 1e-4, 1e-5, 1e-6\}$ & Initial learning rate used by the optimization algorithm. \\ \midrule
            Learning rate scheduler & $\textrm{ExponentialDecay}$ & --- \\ \midrule
            Activation function & $x \in \{\textrm{`relu'},$ $\textrm{`selu'}, \textrm{`gelu'}\}$    & Activation function for the neurons in Dense layers. \\ \midrule
            Dropout rate & $x \in \{0, 0.2, 0.5, 0.7\}$ & Dropout rate. \\ \midrule
            N & $0 \leq x \leq 4$ & Number of feed-forwarded sub-blocks for the first part. \\ \midrule
            M & $0 \leq x \leq 4$ & Number of feed-forwarded sub-blocks for the second part. \\ \bottomrule
        \end{tabular}
    \end{table}
    
    \begin{table}[]
        \centering
        \caption{Best hyperparameter set found after the HPO process for the final deep learning model with a total of $66.7$ million trainable parameters. Version that normalizes input data with Z-score.}
        \label{tab:hpo_nn_best_Z}
        \begin{tabular}{@{}ll@{}}
            \toprule
            \textbf{Hyperparameter} & \textbf{Value} \\ \toprule
            Batch size & $512$ \\ \midrule
            Optimization algorithm & $\textrm{Adam}$ \\ \midrule
            Initial learning rate & $1e-06$ \\ \midrule
            Learning rate scheduler & $\textrm{ExponentialDecay(0.9, 10000)}$ \\ \midrule
            Activation function & $\textrm{relu}$ \\ \midrule
            Dropout rate & $0.5$ \\ \midrule
            N & $4$ \\ \midrule
            M & $0$ \\ \bottomrule
        \end{tabular}
    \end{table}
    
    \begin{table}[]
        \centering
        \caption{Best hyperparameter set found after the HPO process for the final deep learning model with a total of $66.7$ million trainable parameters. Version that does not normalize input data.}
        \label{tab:hpo_nn_best_noZ}
        \begin{tabular}{@{}ll@{}}
            \toprule
            \textbf{Hyperparameter} & \textbf{Value} \\ \toprule
            Batch size & $1024$ \\ \midrule
            Optimization algorithm & $\textrm{Adam}$ \\ \midrule
            Initial learning rate & $1e-05$ \\ \midrule
            Learning rate scheduler & $\textrm{ExponentialDecay(0.9, 10000)}$ \\ \midrule
            Activation function & $\textrm{selu}$ \\ \midrule
            Dropout rate & $0.7$ \\ \midrule
            N & $0$ \\ \midrule
            M & $4$ \\ \bottomrule
        \end{tabular}
    \end{table}

    \subsection{Activity modelling results} \label{sec:activity-experiments}
    
\begin{figure}
        \centering
        \includegraphics[width=0.75\textwidth]{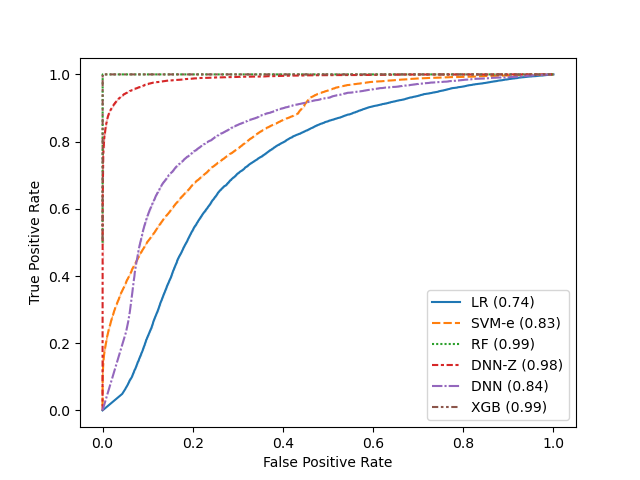}
        \caption{\label{fig:roc_auc__train}Receiver Operating Characteristic (ROC) showing the area under the curve (AUC) for every modeling approach when dealing with the training data partition (A.1). Algorithm name abbreviations: LR (logistic regression), SVM-e (Support vector machines using an ensemble approach), RF (random forests), DNN-Z (deep neural network using Z-score normalization on input data), DNN (deep neural network without Z-score normalization), and XGB (gradient boosting using XGBoost).}
    \end{figure}

\begin{figure}
        \centering
        \includegraphics[width=0.75\textwidth]{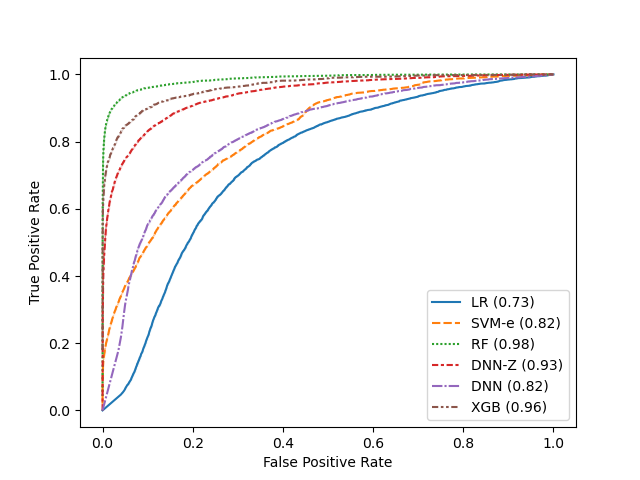}
        \caption{\label{fig:roc_auc__test_ligands}Receiver Operating Characteristic (ROC) showing the area under the curve (AUC) for every modeling approach when dealing with the testing data partition (A.2). Algorithm name abbreviations: LR (logistic regression), SVM-e (Support vector machines using an ensemble approach), RF (random forests), DNN-Z (deep neural network using Z-score normalization on input data), DNN (deep neural network without Z-score normalization), and XGB (gradient boosting using XGBoost).}
    \end{figure}

\begin{figure}
        \centering
        \includegraphics[width=0.75\textwidth]{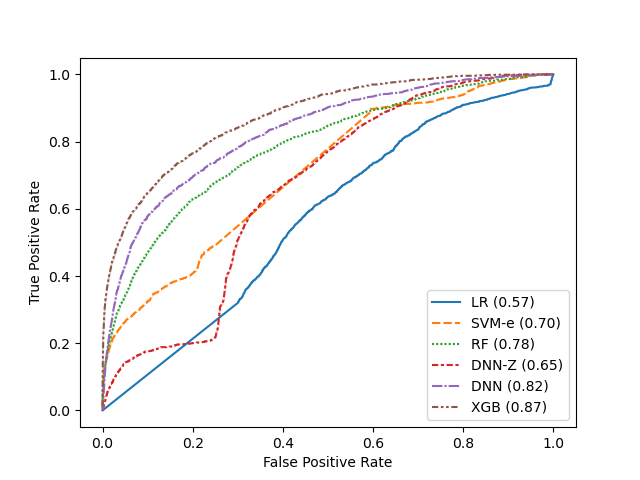}
        \caption{\label{fig:roc_auc__test_ligands_targets}Receiver Operating Characteristic (ROC) showing the area under the curve (AUC) for every modeling approach when dealing with the testing data partition (B). Algorithm name abbreviations: LR (logistic regression), SVM-e (Support vector machines using an ensemble approach), RF (random forests), DNN-Z (deep neural network using Z-score normalization on input data), DNN (deep neural network without Z-score normalization), and XGB (gradient boosting using XGBoost).}
    \end{figure}

    Results in terms of ROC-AUC are summarized in Table \ref{tab:auc_activity} for all the modeling approaches and data partitions. In addition, Figures \ref{fig:roc_auc__train}, \ref{fig:roc_auc__test_ligands}, and \ref{fig:roc_auc__test_ligands_targets} show the area under the ROC curve for the different data partitions (A.1, A.2, and B) respectively, and include all the modeling approaches.
    
    It can be seen that the gradient boosting (XGB) approach obtains a notable AUC performance in all three data partitions: 0.99 in A.1 (training), 0.96 in A.2 (testing with new ligands), and 0.87 in B (testing with new targets and new ligands). Even though the random forest (RF) model outperforms in the A.2 partition obtaining higher AUC (RF=0.98 vs. XGB=0.96), the RF AUC performance in the B partition is worse than XGB (RF=0.78 vs. XGB=0.87). We see this situation as most favorable to the XGB model as both obtain extremely high AUC values in the A.2 partition. However, the performance for XGB in the B partition is higher than RF. We consider the performance in this B partition extremely important as the models face both new target molecules and new ligand molecules not seen during training. This fact puts the RF model in a good position as an alternative, which could be expected as both modeling approaches share some algorithmic principles and how classification decision boundaries are shaped.
    
    However, applying a basic approach such as logistic regression (LR) results in much more modest AUC values. This result is not only due to the linearity constraint on the model decision boundary, where the AUC obtained in the B partition is still much worse than the AUC obtained either in A.1 or A.2. The reason behind this is probably the necessary data preprocessing applied in its training data and later being propagated to the prediction time for filling missing data gaps or data scaling.
    
    The ensemble of support vector machines (SVM-e) obtains slightly better results than the LR model because SVM-e is able to shape classification decision boundaries with non-linearities. Even so, it suffers the same problem as the LR approach, and the AUC performance on the B partition ---with new targets and ligand compounds--- is degraded, probably due to the required data preprocessing steps.
    
    Finally, the deep neural network (DNN) model has been trained with two data preprocessing procedures: one (DNN) with just the basic steps, such as data imputation to fill gaps with missing data, and the other one (DNN-Z) including the data scaling using Z-normalization in addition. The goal is to evaluate the performance impact of data scaling, especially on the testing data partitions. Results show that DNN-Z obtains higher AUC on the training partition A.1 (DNN-Z=0.98 vs. DNN=0.84) and the testing partition A.2 with only new ligands (DNN-Z=0.93 vs. DNN=0.82). However, its performance worsens noticeably on the testing partition B with both new targets and ligands (DNN-Z=0.65 vs. DNN=0.82). The reason is probably the domain shift over the statistics used to scale the new unseen target and ligand compounds. An interesting fact is the performance stability over the three data partitions for the DNN approach, but below its competitors.
    
    It can be concluded that the XGB approach provides the best performance in all the designed data partitions.
    
    \begin{table}[]
        \centering
        \caption{ROC AUC (\textit{Area under the receiver operating characteristic curve}) for the different models and evaluated on the three data partitions. Algorithm name abbreviations: LR (logistic regression), SVM-e (Support vector machines using an ensemble approach), RF (random forests), DNN-Z (deep neural network using Z-score normalization on input data), DNN (deep neural network without Z-score normalization), and XGB (gradient boosting using XGBoost).}
        \label{tab:auc_activity}
        \begin{tabular}{@{}cccc@{}}
            \toprule
            AUC       & \multicolumn{3}{c}{Data partition} \\ \midrule
            Model     & A.1        & A.2        & B        \\ \toprule
            LR        & 0.74073    & 0.73816    & 0.57002  \\
            SVM-e     & 0.83517    & 0.82335    & 0.70706  \\
            RF        & 0.99958    & 0.98542    & 0.78419  \\
            DNN-Z     & 0.98848    & 0.93944    & 0.65947  \\
            DNN       & 0.84338    & 0.82481    & 0.82999  \\
            XGB       & 0.99933    & 0.96384    & 0.87539  \\ \bottomrule
        \end{tabular}
    \end{table}
    
    \subsection{Sensitivity analysis for molecular conformations} \label{sec:sensitivity-analysis}
    
    As noted in Subsection \ref{sec:data-preparation}, certain data preprocessing steps have been performed on the molecular descriptors, including reducing multiple conformations per molecule to a single conformational representative by taking the average of the multiple conformation values. Considering that the $N$ conformations of a given molecule have been generated by rotating each torsion angle $\theta$ degrees in $[0, 360]$, this sample mean is a robust and unbiased estimate from all the generated conformations.
    
    Therefore, rather than viewing this as a potential weakness in information loss, this step is important and useful for adding robustness to estimation and optimization because it reduces the risk of frequency bias. However, it is important to validate this approach during the model inference using all the conformations in order to assess if there is a potential loss of quality in the response.
    
    In this section, we use the XGB model trained for the experiments (Table \ref{tab:hpo_xgb_best}), i.e., trained with the conformation reduction approach. Then, the model is used to perform the inference on the testing data partition B, but now without reducing the conformations to a single representative sample. Therefore, the samples in the dataset correspond to the Cartesian product of all the conformations between all the target proteins and all the ligand compounds.
    
\begin{figure}
        \centering
        \includegraphics[width=0.75\textwidth]{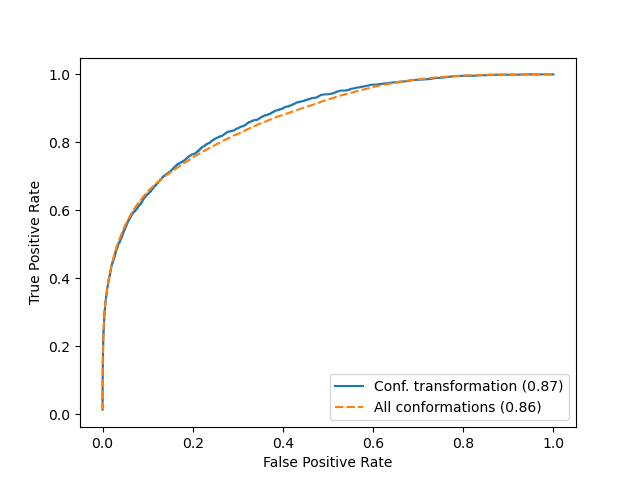}
        \caption{\label{fig:roc_auc__test_ligands_targets__all_conf}Receiver Operating Characteristic (ROC) showing the area under the curve (AUC) for every modeling approach when dealing with the testing data partition (B). Comparison of test prediction data with the conformation transformation versus all the conformations, and using the same model built with XGB (gradient boosting using XGBoost) in both cases.}
    \end{figure}

    Figure \ref{fig:roc_auc__test_ligands_targets__all_conf} shows the ROC AUC results for the prediction over the same testing data partition B both by reducing the conformations and by performing the prediction for all combinations. It can be seen that both curves are almost identical (ROC AUC 0.87 vs. 0.86), so no performance loss is appreciated.
    
\begin{figure}
        \centering
        \includegraphics[width=0.67\textwidth]{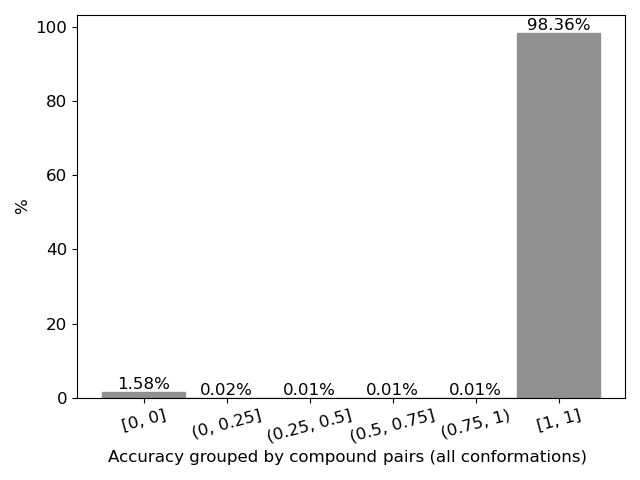}
        \caption{\label{fig:accuracy_mol2mol_allconf}Accuracy is grouped by compound pairs from the testing data partition (B). Each group corresponds to a pair $(a, b)$ containing the Cartesian product of the conformations from a target protein $a$ and a ligand compound $b$. Results show that the model response is robust and consistent as $99.95\,\%$ of the groups (molecule pairs) are classified with a single category for all the conformations, which is why the accuracy is either 0 or 1. Indeed $98.36\,\%$ is 1 ---1 stands for perfect accuracy---.}
    \end{figure}

    Another interesting analysis is checking the performance from a more detailed perspective by calculating the accuracy grouped by compound pairs from the testing data partition (B). Figure \ref{fig:accuracy_mol2mol_allconf} indicates that $99.95\,\%$ of the groups (molecule pairs) are classified into a single category for all the conformations, explaining the reason for obtaining an accuracy of either 0 or 1. 
    
    Indeed  $98.36\,\%$ is 1 ---1 stands for perfect accuracy---. Each group corresponds to a pair $(a, b)$ containing the Cartesian product of the conformations from a target protein $a$ and a ligand compound $b$.
    In conclusion, the results show that the model's response is not highly sensitive to the conformations presented to it, offering a robust, consistent, and quality response.

    \subsection{Molecular similarity results} \label{sec:similarity}
    
    For testing the molecular similarity operation mode, two compounds have been selected from the testing data partition B (see Section \ref{sec:experiments-setup}). One is a small compound with ChEMBL ID \texttt{CHEMBL190}, and the other is a medium-sized compound with ChEMBL ID \texttt{CHEMBL71007}. However, all DUD-E database has been used as search space to find the top 5 most similar compounds.
    
    As discussed in the introduction (Section \ref{sec:intro}), this is a subjective task as there is no single way to determine the similarity between compounds. While some authors may consider the compounds' shape to determine their similarity, others may consider other criteria.
    
    For the similarity operation mode within the ALMERIA methodology presented in this work, two compounds (two ligands) are presented to the trained model that gives a response in $[0, 1]$ describing their pairwise similarity. Here the similarity numbers have been multiplied by 100 simply for easy reading. An advantage of the current ALMERIA methodology and chosen modeling approach is that many queries may be enclosed in a single batch to be resolved by the model at once, thus offering a very quick response. 
    
    Despite the subjectivity, reasonable similarities between the first results can be appreciated. Results with the top-5 most similar compounds are shown in Table \ref{tab:similarity_small_mol} for the small-sized compound and Table \ref{tab:similarity_medium_mol} for the medium-sized compound. Furthermore, to validate our proposal, it is very important to remark that the first most similar compound is always the query compound with $100\%$ similarity.

    \begin{longtable}{cc}
        \caption{Top-5 similarity found in DUD-E for the small-sized compound CHEMBL190 (Molecular weight = $180.16$)}
        \label{tab:similarity_small_mol}
        \endfirsthead
        \endhead
            \toprule
            \thead{ChEMBL ID \\ Molecular formulae \\ Molecular weight \\ Similarity [\%]} & 2D Structure \\
            \midrule
            \makecell{CHEMBL190 \\ (\ch{C7H8N4O2}) \\\\ MW: $180.16$ \\\\ $\textbf{100.0}$}
            &
            \adjustbox{valign=c}{\includegraphics[width=4cm, height=4cm]{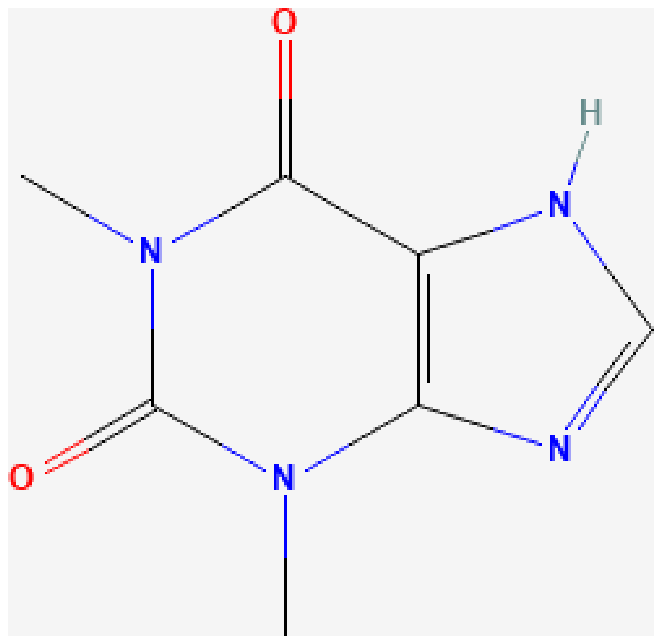}}
            \\
            \midrule
            \makecell{CHEMBL106265 \\ (\ch{C12H16N4O2}) \\\\ MW: $248.28$ \\\\ $\textbf{99.10}$}
            &
            \adjustbox{valign=c}{\includegraphics[width=7cm, height=4cm]{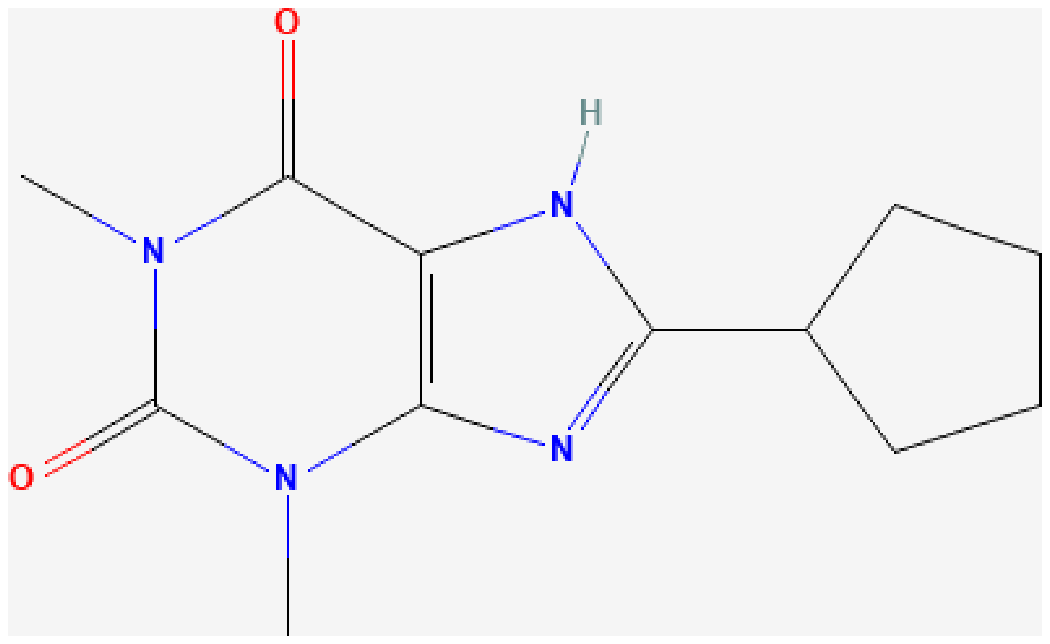}}
            \\
            \midrule
            \makecell{CHEMBL321505 \\ (\ch{C13H18N4O2}) \\\\ MW: $262.31$ \\\\ $\textbf{98.62}$}
            &
            \adjustbox{valign=c}{\includegraphics[width=7cm, height=4cm]{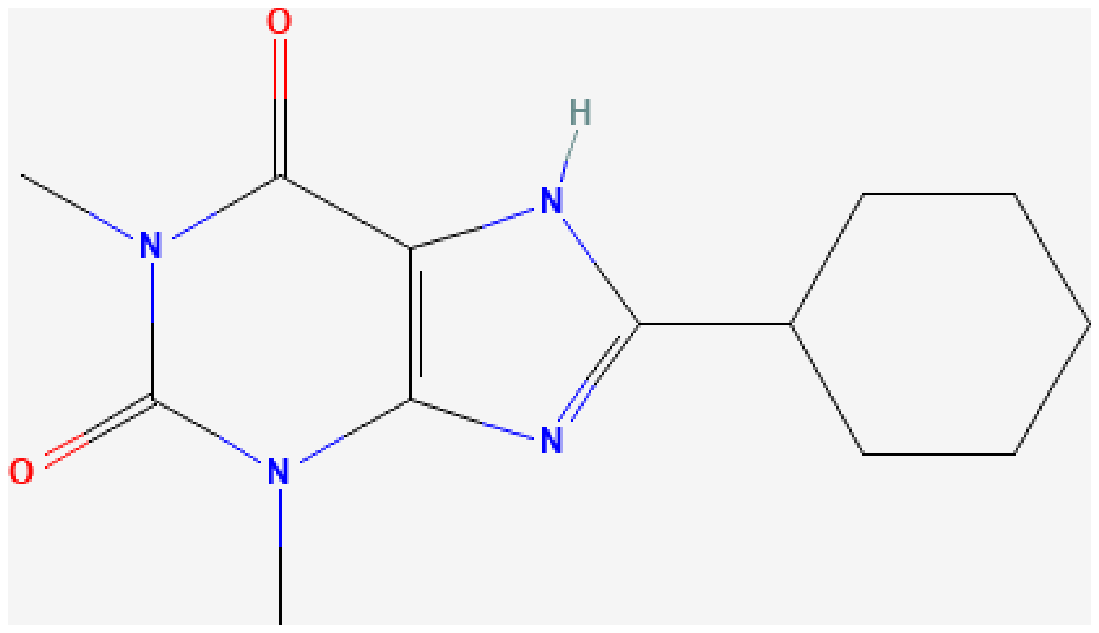}}
            \\
            \midrule
            \makecell{CHEMBL281811 \\ (\ch{C14H20N4O2}) \\\\ MW: $276.33$ \\\\ $\textbf{90.91}$}
            &
            \adjustbox{valign=c}{\includegraphics[width=7cm, height=4cm]{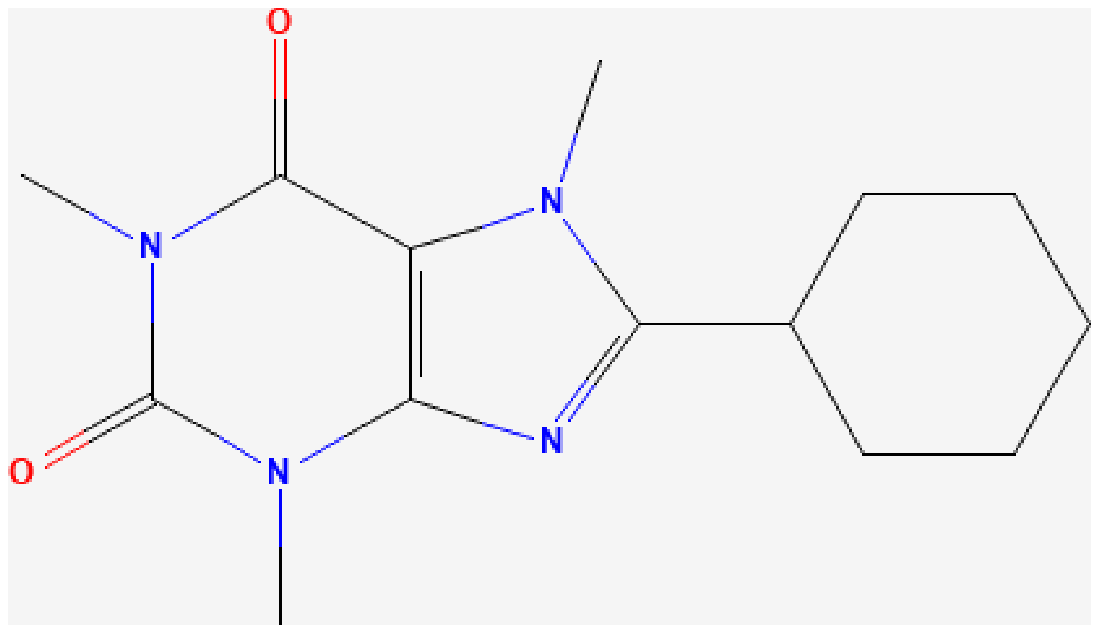}}
            \\
            \midrule
            \makecell{CHEMBL283939 \\ (\ch{C26H28N4O5}) \\\\ MW: $476.5$ \\\\ $\textbf{45.19}$}
            &
            \adjustbox{valign=c}{\includegraphics[width=9.5cm, height=6.5cm]{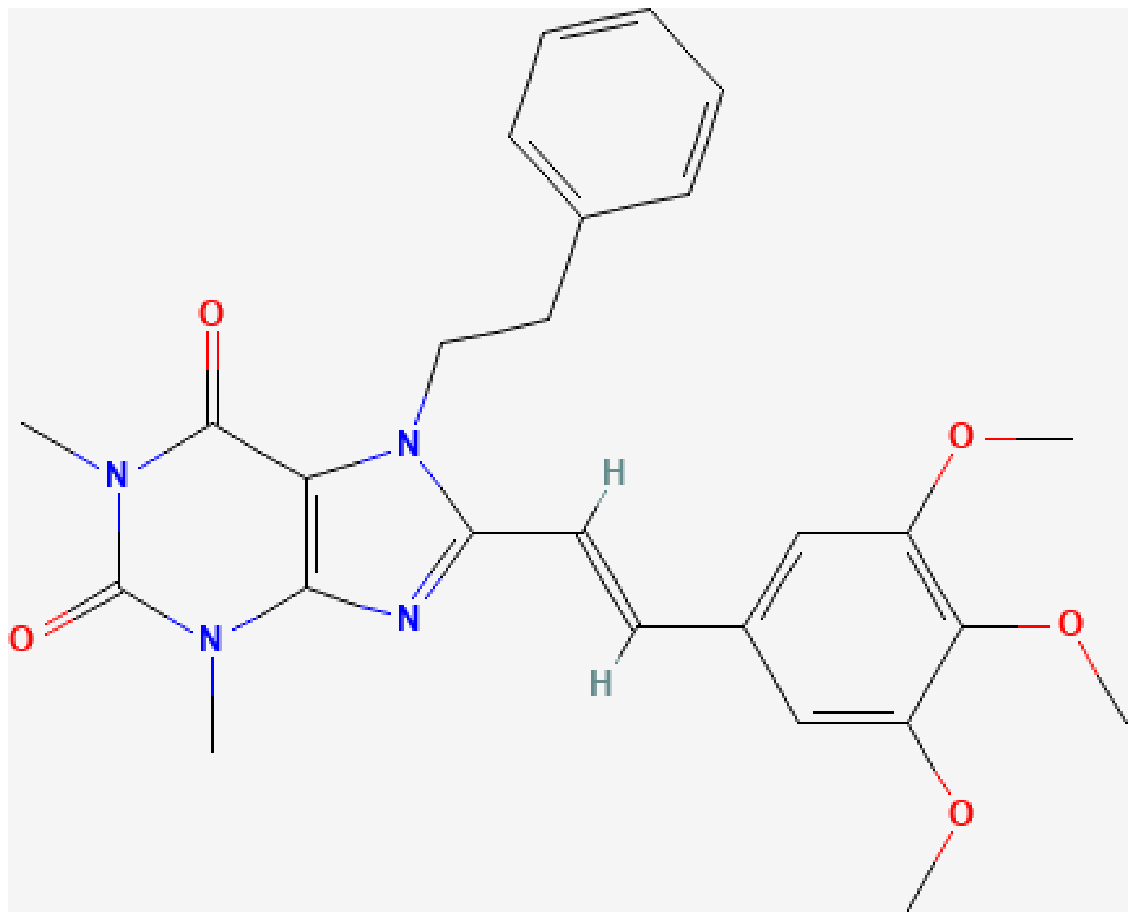}}
            \\
            \bottomrule
    \end{longtable}
    
    \begin{longtable}{cc}
        \caption{Top-5 similarity found in DUD-E for the medium-sized compound CHEMBL71007 (Molecular weight = $530.6$)}
        \label{tab:similarity_medium_mol}
        \endfirsthead
        \endhead
            \toprule
            \thead{ChEMBL ID \\ Molecular formulae \\ Molecular weight \\ Similarity [\%]} & 2D Structure \\
            \midrule
            \makecell{CHEMBL71007 \\ (\ch{C30H30N2O5S}) \\\\ MW: $530.6$ \\\\ $\textbf{100.0}$}
            &
            \adjustbox{valign=c}{\includegraphics[width=10.5cm, height=7cm]{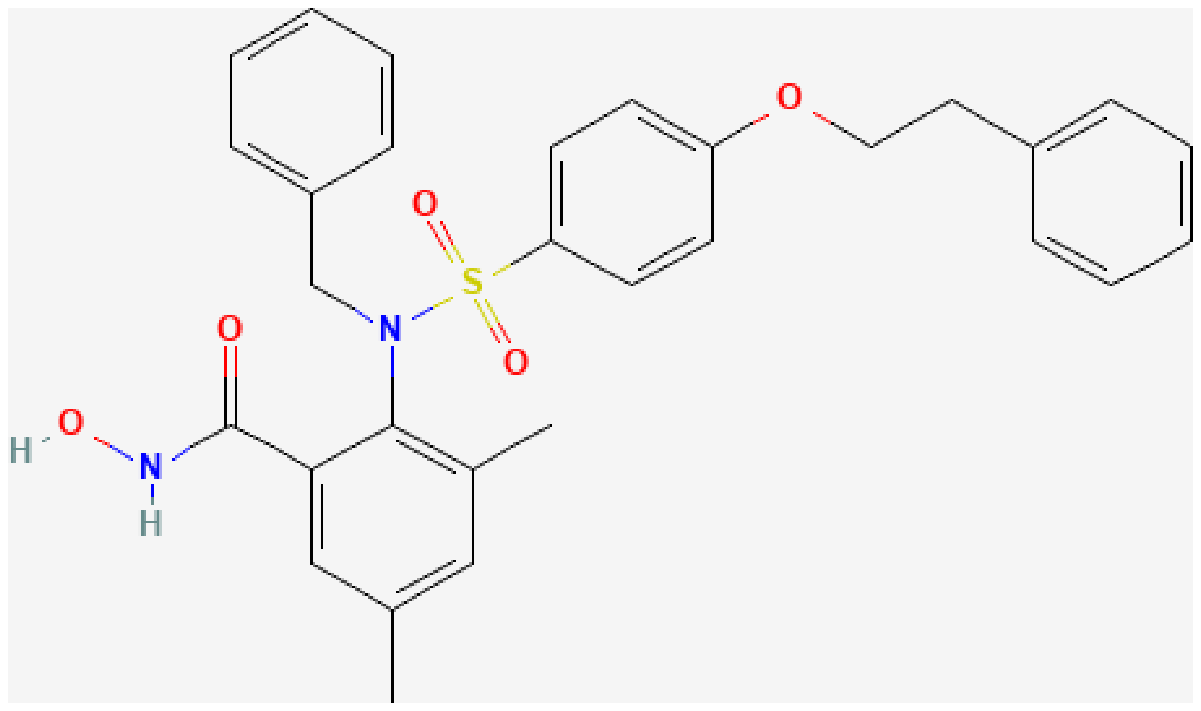}}
            \\
            \midrule
            \makecell{CHEMBL70207 \\ (\ch{C29H28N2O5S}) \\\\ MW: $516.6$ \\\\ $\textbf{100.0}$}
            &
            \adjustbox{valign=c}{\includegraphics[width=10.5cm, height=7cm]{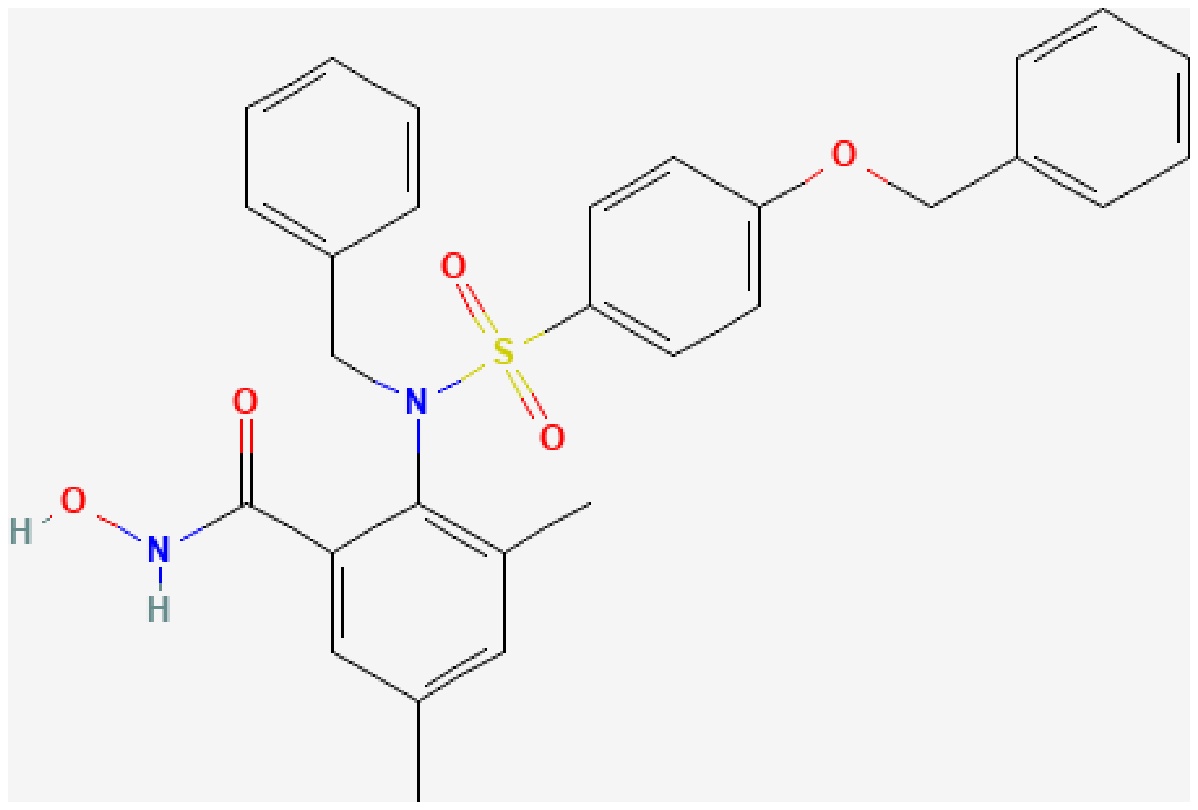}}
            \\
            \midrule
            \makecell{CHEMBL100081 \\ (\ch{C29H35N3O6S}) \\\\ MW: $553.7$ \\\\ $\textbf{99.97}$}
            &
            \adjustbox{valign=c}{\includegraphics[width=8.5cm, height=9cm]{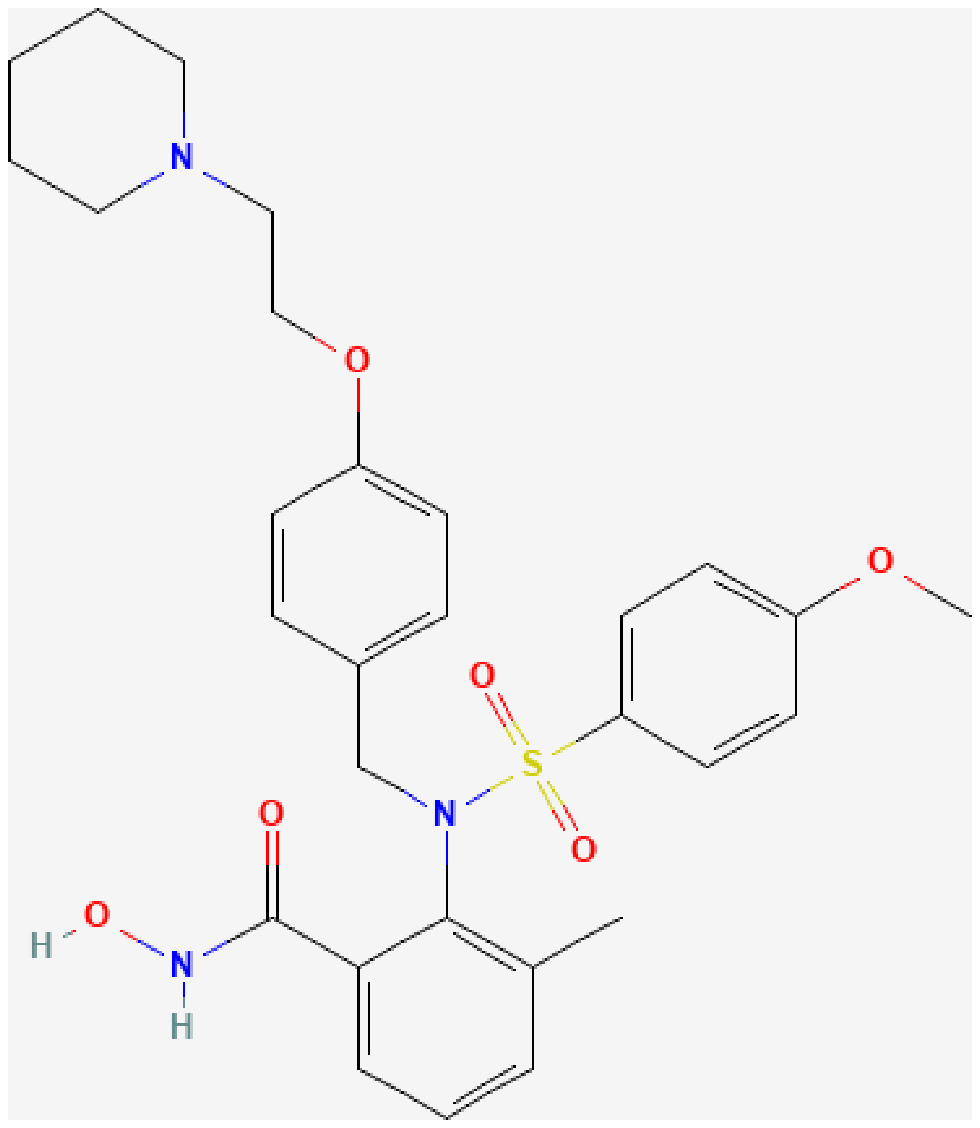}}
            \\
            \midrule
            \makecell{CHEMBL303666 \\ (\ch{C27H25N3O5S}) \\\\ MW: $503.6$ \\\\ $\textbf{99.68}$}
            &
            \adjustbox{valign=c}{\includegraphics[width=8.5cm, height=9cm]{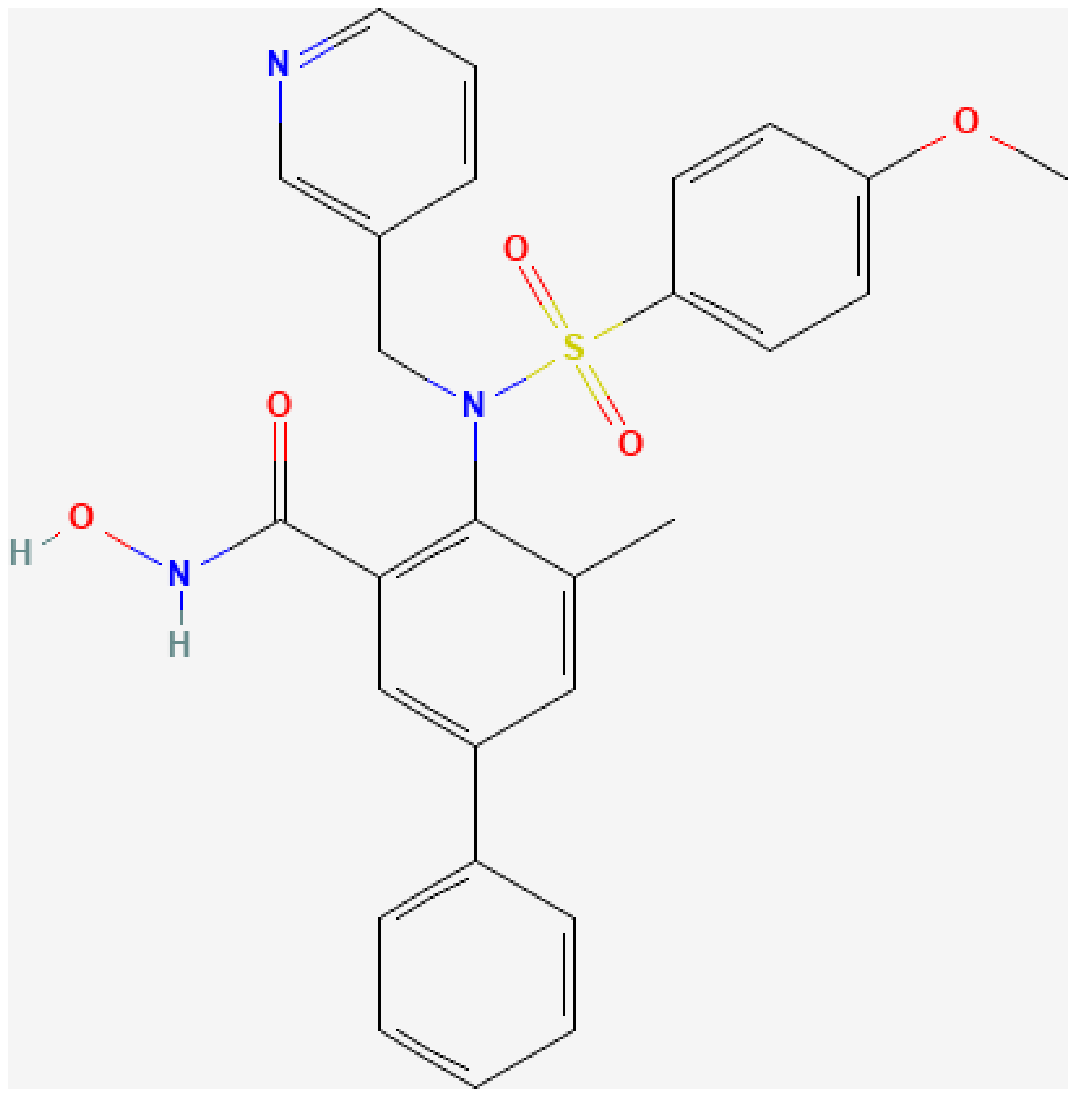}}
            \\
            \midrule
            \makecell{CHEMBL68271 \\ (\ch{C27H25N3O5S}) \\\\ MW: $503.6$ \\\\ $\textbf{99.35}$}
            &
            \adjustbox{valign=c}{\includegraphics[width=8.5cm, height=9cm]{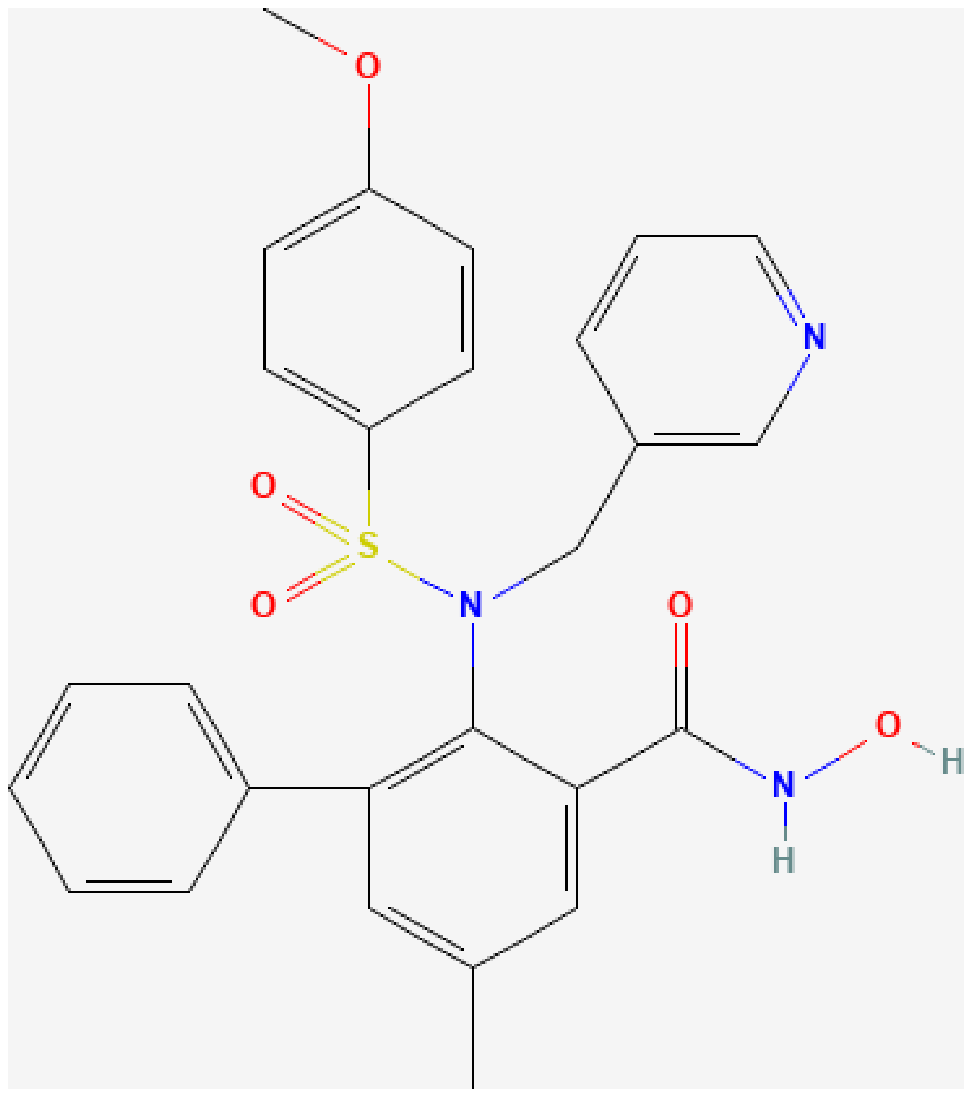}}
            \\
            \bottomrule
    \end{longtable}
    
    
    \section{Model interpretation} \label{sec:interpertability}
    
    We have performed a brief interpretability analysis. For this purpose, we have used the XGB model and have estimated the SHapley Additive exPlanation (SHAP) values over the testing data partition B. 
    SHAP is a common methodology for explainability and interpretability in machine learning. The reader may find more information in the original paper \cite{lundberg_unified_2017} and the detailed optimizations related to tree-shaped machine learning models \cite{lundberg_local_2020}.
    
    First, it is important to note that SHAP values indicate the correlation but not causation. So, all effects and feature importances are framed within the model's behavior and are not necessarily causal in the underlying chemistry. This is especially the case here, where there is a vast number of features, and these are not independent and unconfounded. Nevertheless, performing this interpretability analysis is interesting and may be useful to diagnose the model and check the potential reproducibility of the most important features.
    
    We focus here on global interpretability analysis, even though local interpretability analysis can be performed, for example, by checking the most important effects for a given positive biological activity prediction.
    
    \subsection{Feature importance}
    
\begin{figure}
        \centering
        \includegraphics[width=0.75\textwidth]{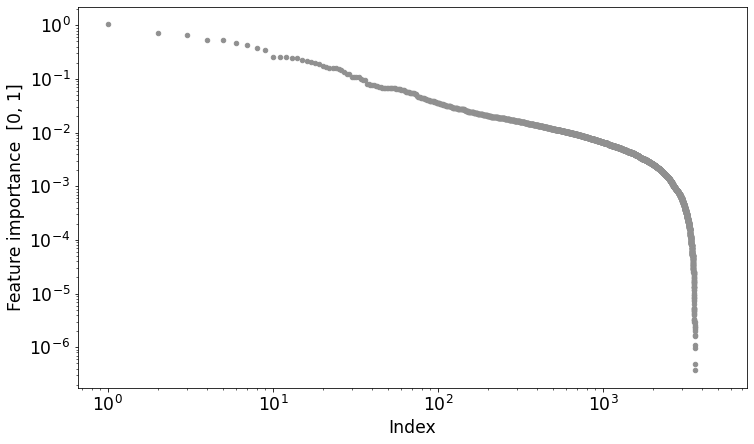}
        \caption{\label{fig:feature_importance_loglog}Feature importance values ---using the SHAP values from the XGB model over the testing data partition B--- in decreasing order and logarithmic scale.}
    \end{figure}

\begin{figure}
        \centering
        \includegraphics[width=0.75\textwidth]{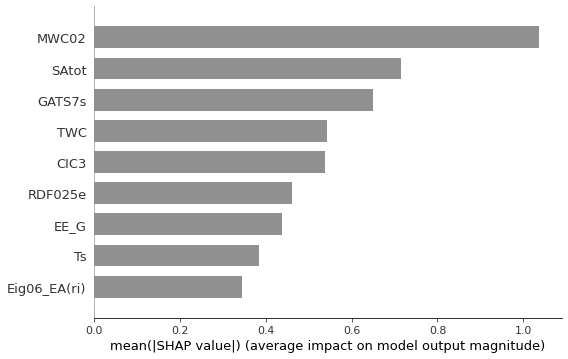}
        \caption{\label{fig:feature_importance_top9}Top 9 features according to their estimated higher importance using the SHAP values.}
    \end{figure}

    The first part is to check for the most important features among the $4\,885$ numerical descriptors. However, basing the decision on an arbitrary number of features or an arbitrary threshold of feature importance is not trivial. To this end, we have observed the feature importance values in decreasing order and using a logarithmic scale as shown in Figure \ref{fig:feature_importance_loglog}. This way, it can be more easily identified in the chart's upper left corner as a cluster with nine features whose importance is the highest. From the tenth onwards, the importance is around $10^{-1}$ and decreasing, but it is more important the fact that the relative difference among features' importance values become much more compressed ---which is also observed in a linear scale---. For this reason, the first nine features, as shown in Figure \ref{fig:feature_importance_top9}, have been selected for interpretability. The names for these features and their parent descriptor category can be seen in Table \ref{tab:important_features_names}.
    
    \begin{table}[]
        \centering
        \caption{Top 9 features according to their estimated higher importance using the SHAP values.}
        \label{tab:important_features_names}
        \begin{tabular}{@{}lp{7cm}l@{}}
            \toprule
            Feature abbr. & Feature name & Block \\ \midrule
            MWC02 & Molecular walk count of order 2 & Walk and path counts \\
            SAtot & Total surface area from P\_VSA-like descriptors & Molecular properties \\
            GATS7s & Geary autocorrelation of lag 7 weighted by I-state & 2D autocorrelations \\
            TWC & Total walk count & Walk and path counts \\
            CIC3 & Complementary Information Content index (neighborhood symmetry of 3-order) & Information indices \\
            RDF025e & Radial Distribution Function - 025 / weighted by Sanderson electronegativity & RDF descriptors \\
            EE\_G & Estrada-like index (log function) from geometrical matrix & 3D matrix-based descriptors \\
            Ts & T total size index / weighted by I-state & WHIM descriptors \\
            Eig06\_EA(ri) & Eigenvalue n. 6 from edge adjacency mat. weighted by resonance integral & Edge adjacency indices    
        \end{tabular}
    \end{table}
    
    \subsubsection{Feature importance per target protein}
    
    A more detailed analysis has also been carried out to identify the feature importance of individual target proteins. These proteins exist in the testing data partition B (see Subsection \ref{sec:experiments-setup}). 
    
    The five most important descriptors identified are shown in Table \ref{tab:important_features_groupby_target}. The results show that, although the order for each protein does not coincide exactly with that previously obtained for the whole dataset, the frequency of the most important ones tends to repeat itself among the top 5.
    
    \begin{table}[]
        \centering
        \caption{Top 5 most important features according to the estimated SHAP values and grouped by target protein. The left-most column is the most important one. Target proteins correspond to the testing data partition B.}
        \label{tab:important_features_groupby_target}
        \begin{tabular}{@{}clllll@{}}
            \toprule
            Target protein & \multicolumn{5}{l}{Top-5 feature importance} \\ \midrule
            aa2ar          & MWC02  & SAtot  & GATS7s & CIC3    & TWC     \\
            abl1           & SAtot  & MWC02  & CIC3   & GATS7s  & NsCl    \\
            ace            & MWC02  & SAtot  & GATS7s & CIC3    & RDF025e \\
            ada17          & MWC02  & GATS7s & SAtot  & GATS4e  & TWC     \\
            akt1           & MWC02  & CIC3   & SAtot  & RDF025e & TWC     \\
            andr           & MWC02  & SAtot  & TWC    & LDI     & CIC3    \\ \bottomrule
        \end{tabular}
    \end{table}
    
    \subsection{Feature effect}
    
\begin{figure}
        \centering
        \includegraphics[width=0.75\textwidth]{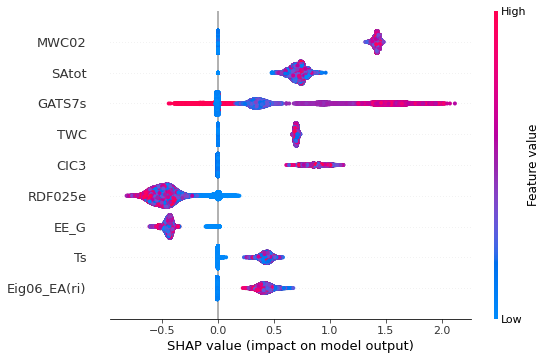}
        \caption{\label{fig:feature_effects_top9}Top 9 features according to their estimated higher importance using the SHAP values. Each point per feature is a SHAP value for a given data sample so that SHAP data density may be perceived. The chart shows the relationship between the value of a specific feature (colored low to high) and its impact on the prediction measured by the magnitude of the SHAP value.}
    \end{figure}

    These selected features (Figure \ref{fig:feature_importance_top9}) can be analyzed in more detail by combining their importance ---decreasing order along the y-axis--- and their effect. In Figure \ref{fig:feature_effects_top9}, each point per feature is a SHAP value for a given data sample ---which is the reason why overlapping points are jittered so SHAP data density may be perceived---. This chart shows the relationship between the value of a specific feature ---colored low to high--- and its impact on the prediction ---measured by the magnitude of the SHAP value---. 
    
    For example, in the case of the feature identified as the most important one (\texttt{MWC02}), it is clear that values of zero have a zero impact, while larger values have a significant (positive) impact on the prediction. 
    
    However, features \texttt{RDF025e} and \texttt{EE\_G} show similar behavior. Contrarily, on the opposite side, since, in this case, the impact on predictions is negative, i.e., they attract the instance to a probability of zero, potentially labeling the instance as not biologically active. 
    
    \textbf{An important reminder:} Finally, it is important to remember that due to the transformation in the representation of the problem  (see Section \ref{sec:data-preparation}), the values of the features do not reflect their natural values but their relative difference between a pair of compounds, which implies that the effect of the feature explains this new representation.
    
    \subsection{The interaction power of the features effect}
    
    Previously in Figure \ref{fig:feature_effects_top9}, some first indications of the relationship between the value of a feature and the impact on the prediction were shown. However, the specific shape of this relationship may be observed through a SHAP dependence plot. 
    
\begin{figure}
        \centering
        \includegraphics[width=0.75\textwidth]{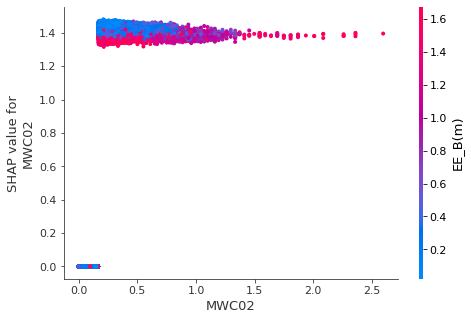}
        \caption{\label{fig:feature_dependence_rank0}Feature interaction between values from the variable \texttt{MWC02} (identified as the most important one in the model) on the x-axis and the variable \texttt{EE\_B(m)} colored from low to high. Each dot is a single prediction (data sample) from the dataset. The y-axis is the SHAP value standing for the impact on the prediction.}
    \end{figure}

    For example, Figure \ref{fig:feature_dependence_rank0} shows the interaction between values from the variable \texttt{MWC02}, which was identified previously as the most important one in the model, on the x-axis against the variable \texttt{EE\_B(m)}. This variable \texttt{EE\_B(m)}  stands for \textit{Estrada-like index (log function) from Burden matrix weighted by mass}, a 2D matrix-based descriptor. This Figure \ref{fig:feature_dependence_rank0} is colored from low to high, where each dot is again a single prediction (data sample) from the dataset. The y-axis is the SHAP value standing for the impact on the prediction. In this case, it can be seen that low values on the x-axis (\texttt{MWC02}) are tied to a null impact on the prediction independently of the interaction color (\texttt{EE\_B(m)}). However, beyond a certain threshold value of \texttt{MWC02} (around $0.25$), it may be noted that lower values of \texttt{EE\_B(m)} have a greater impact on the prediction (greater SHAP value on the y-axis). The highest values of \texttt{MWC02} are only related to high values of \texttt{EE\_B(m)}.
    
\begin{figure}
        \centering
        \includegraphics[width=0.75\textwidth]{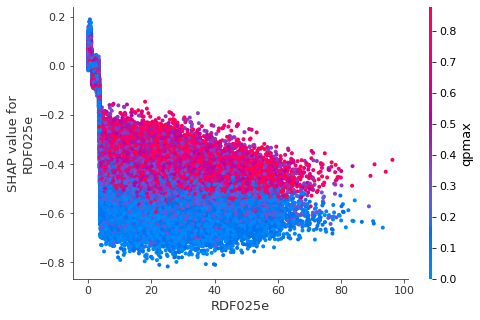}
        \caption{\label{fig:feature_dependence_rank5}Feature interaction, colored from low to high, between values from the variable \texttt{RDF025e} (identified as the 6th most important one in the model) on the x-axis and the variable \texttt{qpmax}. Each dot is a single prediction (data sample) from the dataset. The y-axis is the SHAP value standing for the impact on the prediction.}
    \end{figure}

    Another example is in Figure \ref{fig:feature_dependence_rank5}, where the interaction is shown, colored from low to high, between values from the variable \texttt{RDF025e}, which was identified previously as the 6th most important one in the model, on the x-axis against the variable \texttt{qpmax}. This variable \texttt{qpmax} stands for \textit{maximum positive charge}, a charge descriptor. The rest is the same as the previous chart: each dot is again a single prediction (data sample) from the dataset, and the y-axis is the SHAP value standing for the impact on the prediction. For this case, very low values of \texttt{RDF025e} result in a slightly positive effect over the prediction independently of the \texttt{qpmax} value. However, as the value of \texttt{RDF025e} is higher than a certain threshold, the effect over the prediction turns out to be negative. The effect appears to become even more negative-biased as the \texttt{qpmax} value is low. 
    
    These have been only some examples, and the analysis could be more comprehensive. However, it is beyond the scope of the current work, and we will leave it for future work in combination with more experts in the field of chemistry.
    
    \section{Performance measurement} \label{sec:perf-measurement}
    
    In order to measure the performance in terms of computing times for the different steps in the process (see Figure \ref{fig:pipeline}), only one node of the cluster has been used: 
    
    \begin{itemize}
        \item For CPU computing mode: AMD EPYC Rome 7642 (48 cores) with 512 GB RAM and 240 GB SSD. 
        \item For GPU computing mode: AMD EPYC Rome 7302 (16 cores) with 512 GB RAM and 240 GB SSD, 2x GPUs NVIDIA Tesla V100 with 32GB HBM2, 5\,120 CUDA cores, and 640 Tensor cores.
    \end{itemize}
    
    However, the shared network file system (NFS) has been used for disk input/output operations. It has significant repercussions on the first steps of the pipeline concerning the generation of conformations and descriptors.
    
    A single protein target from the DUD-E database has been used to accelerate the performance measurements. The protein target is \texttt{aa2ar}, associated with $32\,909$ compounds ($31\,550$ of which are decoys).
        
    The following computing times have been obtained:
    
    \begin{itemize}
        \item Data preparation
        \begin{itemize}
            \item Generating maximum 100 conformations for each of the $32\,909$ compounds using Omega software: 5 hours 30 minutes. 
            \item Generating molecular descriptors for each conformation from all the $32\,909$ compounds ---$2\,594\,901$ data samples--- using Dragon software: 14 hours.
            \item Data size for this single crystal \texttt{aa2ar} with maximum 100 conformations per compound: $2\,594\,901$ data samples, $4\,885$ columns, $47.27$ GB using 32-bit precision. For the entire DUD-E database: $264\,679\,902$ data samples, $4.7$ TB.
            \item Reading database: 10 minutes.
            \item Reduce conformations to a single representative sample: $<$ 2 minutes.
            \item Compound pair data transformation: $<$ 1 minute.
        \end{itemize}
        \item Model building
        \begin{itemize}
            \item CV folds creation: $<$ 1 minute.
            \item Hyperparameter optimization with 100 trials and using 10-fold CV per trial (CPU): 14 hours.
            \item Hyperparameter optimization with 100 trials and using 10-fold CV per trial (GPU): 6 hours.
            \item Final model training: $<$ 1 minute.
        \end{itemize}
        \item Model inference
        \begin{itemize}
            \item Activity and similarity prediction on full dataset with all compound pairs: $<$ 1 second.
        \end{itemize}
    \end{itemize}
    
    \section{Conclusion} \label{sec:conclusion}
    
    A methodology (ALMERIA: Advanced Ligand Multiconformational Exploration with Robust Interpretable Artificial Intelligence) has been proposed to develop virtual screening software for estimating compound similarity and activity prediction. Its core is based on pairwise molecular contrasts while also considering the conformation variability. The great benefit is obtaining excellent classification rates on out-of-sample observations and having a quick response, even for a large batch of queries. Moreover, this proposal relies on artificial intelligence methods to exploit high-dimensional data instead of over-optimizing based on a specific criterion (e.g., shape). At the same time, efforts have been put on explainable AI (XAI).
    
    As shown in Figure \ref{fig:pipeline} and described in Section \ref{sec:materialsmethods}, the ALMERIA methodology has covered the entire pipeline from data preparation to model selection and hyperparameter optimization. In this work, the implementation is based on scalable software and methods to exploit large volumes of data ---in the order of several terabytes---, and deployed in a distributed computer cluster using a real use case: the public access DUD-E database. However, the ALMERIA methodology is generic enough to be applied to any other database that exploits molecular contrasts.
    
    The chosen data representation is based on numerical molecular descriptors ---e.g., as generated by the Dragon software \cite{mauri_dragon_2006}--- that is generated for each conformation of every molecule. These conformations were generated using OpenEye Scientific Omega software \cite{hawkins_conformer_2010} with a limit of 100 conformations per molecule. Two transformations have been applied to these data representations:
    
    \begin{enumerate}
        \item Reducing multiple conformations for a given molecule to a single representative sample using the averaged descriptors values, thus reducing frequency bias on the model optimization process. Experiments and sensitivity analysis in Section \ref{sec:sensitivity-analysis} show that model response is consistent among multiple combinations of conformation pairs for different molecules. 
        \item Transforming the molecules' descriptors to pairwise molecular contrasts using the absolute difference between their descriptor values. This way, a single model may fit the entire database, therefore enjoying better generalization properties as shown by the experiments in Section \ref{sec:activity-experiments} on numerical molecular descriptors, as the one generated by the Dragon software \cite{mauri_dragon_2006} for each conformation of every molecule. These conformations were generated using OpenEye Scientific Omega software \cite{hawkins_conformer_2010} with a limit of 100 conformations per molecule.
    \end{enumerate}
    
    A very important aspect of the ALMERIA methodology is to have used detailed data split criteria in addition to the cross-validation used during the HPO process. In this way, the models' predictive performance is evaluated on different data partitions to assess their true generalization ability with protein targets and ligands not seen previously during training or validation. The designed data partitions for the use case in this work (DUD-E database) are shown in Section \ref{sec:experiments-setup}.
    
    The underlying machine learning algorithm for similarity and activity prediction is based on a supervised classification model. Any model that satisfies these conditions may be plugged into the proposed pipeline (Figure \ref{fig:pipeline}). Our main proposal is based on gradient boosting after studying the problem characteristics (Section \ref{sec:gradient-boosting}), but other models such as logistic regression, SVM ensemble, random forests, and deep neural net have been included to benchmark the performance. Every model architecture has been optimized using a thorough hyperparameter optimization process using 10-fold cross-validation.
    
    The best molecular activity prediction results are obtained with the main modeling proposal gradient boosting, showing the state-of-the-art performance (ROC AUC: $0.99$, $0.96$, $0.87$), especially with the data partition whose protein targets and ligands are new for the model. This result also proves that the chosen data representation and modeling have good generalizable properties. 
    
    As mentioned above, molecular conformation information is not neglected for every molecule but is used in a new data transformation instead. As this can be controversial, especially in the chemical field, a sensitivity analysis (Section \ref{sec:sensitivity-analysis}) was performed, indicating that the model response is consistent and no performance loss is observed when the model is queried using a Cartesian product of all the molecular conformations.
    
    Moreover, the modeling proposal has demonstrated that it may offer interesting and useful results for compound similarity (Section \ref{sec:similarity}). 
    
    An interpretability analysis using the ALMERIA methodology based on the Shapley Additive Explanations (SHAP) has been carried out in Section \ref{sec:interpertability}. Results show which descriptors are most influential on the model decisions globally and for each protein target. The features' effect and interaction are also presented as examples, showing the ability to diagnose local queries on demand. 
    
    Finally, a small performance measurement exercise has been performed (Section \ref{sec:perf-measurement}) to measure the elapsed time (in CPU or GPU computing mode) for every important step within the ALMERIA methodology pipeline. The model efficiency allows us to respond quickly, even for a large batch of queries.
    
    In future work, we will consider the use of additional molecular databases both for performing the inference with the model trained here and applying the entire methodology with additional merged databases. Moreover, it would be interesting to perform a more in-depth interpretability analysis, even accommodating causality tools and collaboration with additional experts in the field of chemistry.
    
    Although a probability value $[0, 100]$ or similarity score $[0, 1]$ may be easility interpreted by the end-user of the system, we would like to be more rigorous in measuring uncertainty, for example by giving confidence intervals.
    
    Future efforts will be put on letting the system be able to perform online learning, i.e. to update the already learned model with new data as it is collected. This should not be a great challenge as the specific gradient boosting implementation (XGBoost) already allows to perform incremental updates.
    
    \section*{Acknowledgements}
    
    This work has been partially supported by Grant PID2021-123278OB-I00 funded by MCIN/AEI/ 10.13039/501100011033 and by “ERDF A way of making Europe”, and
    by J. Andaluc\'ia through Projects UAL18-TIC-A020-B and P18-RT-1193. 
    
    \bibliography{cheminformatics, ml, hpo} 
    \bibliographystyle{elsarticle-num}
    
\end{document}